\documentclass[a4paper,12pt]{article}
\usepackage{jheppub} % for details on the use of the package, please see the JINST-author-manual
\usepackage{lineno}
\usepackage[numbers,sort&compress]{natbib}
\usepackage{graphicx} % Required for inserting images
\graphicspath{{./images/}}
\usepackage{braket}
\usepackage[boxsize=0.8em,centertableaux]{ytableau}
\usepackage{amsmath, amssymb, amsthm}
\usepackage{braket}
\usepackage{tikz} %使用tikzpicture
\numberwithin{equation}{section}  %公式显示章节号
\usepackage{mdframed}
\usepackage{xcolor} %给文字加颜色
\usepackage{amsfonts}
\usepackage{empheq}
\usepackage{mathrsfs}
\usepackage{mathtools}
\usepackage{subcaption}    % 子图排版
\usepackage{caption}       % 标题样式
\newtheorem{Lemma}{Lemma}

\usepackage[bottom]{footmisc} %强制脚注与引用同页
\title{Charge functions for odd dimensional partitions}

\author[a]{Shang Xiang,}
\author[a]{Hao Feng,}
\author[a]{Keyou Zhuo,}
\author[a]{Tian-Shun Chen,}
\author[a,b,c,1]{and Kilar Zhang\note{Corresponding Author.}}

\affiliation[a]{Department of Physics and Institute for Quantum Science and Technology, Shanghai University, Shanghai 200444, China}
\affiliation[b]{Shanghai Key Lab for Astrophysics, Shanghai 200234, China}
\affiliation[c]{Shanghai Key Laboratory of High Temperature Superconductors, Shanghai 200444, China}

\emailAdd{srythx@shu.edu.cn}
\emailAdd{fenghaozi@shu.edu.cn}
\emailAdd{zhuokeyou@shu.edu.cn}
\emailAdd{cts2003912@shu.edu.cn}
\emailAdd{kilar@shu.edu.cn}

\abstract{
To construct a BPS algebra with representations furnished by n-dimensional partitions, the first step is to find the eigenvalues of the Cartan operators acting on them. The generating function of the eigenvalues is called the charge function. It has an important property that for each partition, the poles of the function correspond to the projection of the boxes which can be added to or removed from the partition legally. The charge functions of lower dimensional partitions, i.e., Young diagrams for 2D, plane partitions for 3D and solid partitions for 4D, are already given in the literature. In this paper, we propose an expression of the charge function for arbitrary odd dimensional partitions and have it proved for 5D case. Some explicit numerical tests for 7D and 9D case are also conducted to confirm our formula.}

\makeatletter
\gdef\@fpheader{} %首页不显示页眉
\makeatother

\begin{document}
\maketitle

\newpage
\section{Introduction}
Partitions, as fundamental combinatorial  objects, have long played an important role in both mathematics and physics. The study of integer partitions -- ways of writing a positive integer as a sum of positive integers -- has deep connections to number theory, representation theory and statistical mechanics~\cite{andrews1998theory}. In 2D, Young diagrams provide a geometric visualization of partitions and arise naturally in the representation theory of symmetric groups and in the geometry of Hilbert schemes of points on surfaces~\cite{nakajima1999lectures,okounkov2007random}. Their generating function is given by the celebrated Euler product formula, and they serve as building blocks for many integrable systems and random matrix models~\cite{mehta2004random,gross1980possible}.

The natural generalization to 3D, known as plane partitions~\cite{macmahon1898memoir}, corresponding to stacking boxes in the positive octant of $\mathbb{Z}^{3}$ subject to non-increasing conditions along three axes. Plane partitions are intimately related to the geometry of $\mathbb{C}^3$, the simplest toric Calabi-Yau threefolds (CY$_{3}$), where they count the equivalent Bogomol'nyi-Prasad-Sommerfield (BPS) states of D6-D0-branes in string theories~\cite{ooguri2009crystal,okounkov2006quantum,iqbal2008quantum}. The generating function of plane partitions is given by the MacMahon function, which also appears in the context of topological string theory and Donaldson-Thomas invariants~\cite{donaldson1998gauge,Kontsevich:2010px,maulik2006gromov,maulik2019quantum}.

For 4D, the analogous objects are solid partitions~\cite{macmahon1912ix}. They can be visualized by placing boxes in the positive corner of a 4D space. Physically, they enumerate equivariant BPS states of D8-D0-branes on the Calabi-Yau fourfolds (CY$_{4}$), $\mathbb{C}^4$, generalizing the correspondence between plane partitions and D6-D0-states on $\mathbb{C}^3$~\cite{nekrasov2019magnificent,bonelli2023adhm}. Solid partitions are notoriously difficult to enumerate and their elusive generating function is believed to encode the partition function of D8-D0-branes on $\mathbb{C}^4$. The conjectured generating function of solid partitions introduced in~\cite{macmahon2004combinatory} failed at level 6, giving 141 instead of the true value 140 (see developments in~\cite{nekrasov2020magnificent,nekrasov2019magnificent,bonelli2023adhm,szabo2023instanton,kimura2023double,nekrasov2024global,cao2018zero}).

A BPS algebra emerges by rendering the space of BPS states with an algebraic structure. Harvey and Moore formulated a generic approach to BPS algebras in terms of scattering in generic supersymmetric systems~\cite{harvey1996algebras,harvey1998algebras} and some applications to D-branes on CY’s were developed in~\cite{Kontsevich:2010px}. The BPS algebra in terms of Chevalley basis with three types of operators: Cartan operators, creation operators and annihilation operators is presented in \cite{li2020quiver}. They can be combined respectively into three generating fields of a complex spectral parameter $u$. For each partition, the eigenvalues of the Cartan generating operators are packaged into a meromorphic function of 
$u$, known as the charge function \cite{li2020quiver,galakhov2024charging}, which in turn governs the action of the creation and annihilation operators. With a certain ansatz on the actions of operators of the BPS algebra on the representation vector labeled by an $n$-dimensional partition, one may determine the BPS algebra with the help of its representation theory. This is introduced as a bootstrap procedure in~\cite{li2020quiver}. Recently, there is an alternative to this approach called double quiver Yangian algebras, where the pole constraint is lifted~\cite{Bao:2025hfu,Bao:2025dqs}. A key property of the charge function, which is essential for the BPS algebra, is that its poles are in one-to-one correspondence with the box positions where a box can be added by a creation operator or removed by an annihilation operator. So to construct a BPS algebra, an attempt to mimic the bootstrap procedure in higher dimensions necessarily begins with the construction of a charge function with such constraints proposed by the alleged bootstrap. 

For plane partitions, this property forces the charge function to assume a factorized form: it is simply a product over all boxes with each contributing a basic rational function which is the core ingredient defining the BPS algebra. This algebra is the affine Yangian of $\mathfrak{gl}_{1}$ describing the BPS states of D6-D0-branes on $\mathbb{C}^3$ with the representation space labeled by plane partitions~\cite{prochazka2016,rapcak2020cohomological,galakhov2024charging,Rapcak:2021hdh}. It is also known to be isomorphic to the central extension of Spherical degenerate double affine Hecke algebra (SH$^c$) introduced by Shffmann and Vasserot in~\cite{schiffmann2013cherednik,arbesfeld2013presentation}, to describe the equivalent cohomology of the instanton moduli space of $\mathcal{N}=2$ 4D gauge theories. This algebra precisely describe the algebraic structure behind Nekrasov instanton partition functions~\cite{nekrasov2003seiberg} with the Omega background and has been used to prove the 4D/2D correspondence proposed by Alday, Gaiotto and Tachikawa (AGT correspondence~\cite{alday2010liouville}) and its various generalization~\cite{kanno2013extended,matsuo2014construction,bourgine2016holomorphic,bourgine2016coherent,bourgine2019note}.

Currently, the corresponding BPS algebra for solid partitions remains unknown, and in~\cite{nekrasov2020magnificent} this novel algebraic structure is termed as \textit{Mama} algebra. In a significant advance, Galakhov and Li illuminatingly resolve the problem of the charge function in 4D case~\cite{galakhov2024charging}.\footnote{A straightforward generalization of the simple form of charge functions in 3D to 4D can not work, since the charge function does not reproduce the correct poles structure associated with the addable and removable boxes for solid partitions.} Some extra contributions from certain 4-box and 5-box clusters are introduced apart from the contributions from single boxes. They prove the constructed charge function satisfies all the required properties for any solid partition by checking all local pictures explicitly.

%The Alday-Gaiotto-Tachikawa (AGT) correspondence \cite{alday2010liouville} \cite{}has fundamentally reshaped our understanding of supersymmetric theories and two-dimensional conformal field theories (CFTs). Inspired by the seminal work on deriving the exact low-energy effective action of four-dimensional $\mathcal{N}=2$ supersymmetric gauge theories from supersymmetry and Riemann surface theory \cite{seiberg1994electric,seiberg1994monopoles}, subsequent developments led to techniques for computing exact partition functions of supersymmetric theories in the $\Omega$-background, known as supersymmetric localization \cite{nekrasov2003seiberg,pestun2012localization}. Furthermore, the scope of studying four-dimensional $\mathcal{N}=2$ supersymmetric theories was greatly expanded through connections with M-theory and Hitchin systems \cite{gaiotto2012n}. These advances laid the groundwork for the AGT correspondence, which posits the exact equivalence between the instanton partition function of a four-dimensional $\mathcal{N}=2$ theory and the conformal blocks of a two-dimensional Toda CFT\cite{}.

In the line of this development, we aim to advance this program of constructing charge functions for higher dimensional partitions which are also meaningful for higher dimensional CY's explained in~\cite{Closset:2018axq,Franco:2019bmx,Franco:2020ijt,Franco:2020avj} Consequently, we conjecture an expression of the charge function for any odd dimensional partition and prove it for 5D case. Instead of using the proof method for the 4D case in \cite{galakhov2024charging}, which relies on an exhaustive classification of local neighborhood constraints in the projection space, our approach employs a global mathematical induction on the total number of boxes. By decomposing the potential into a local hypercube contribution and a background partition, this method circumvents the combinatorial complexity of classifying high-dimensional local configurations, thereby offering a more scalable framework for generalization. We also conduct some explicit numerical tests for 7D and 9D to verify the formula. For higher even dimensional cases, it is solved in a following work \cite{Feng:2025pft}.

This paper is organized as follows. In section \ref{sec:n-dimensional partitions}, we review some necessary concepts about partitions in arbitrary dimension. In section \ref{sec:Conjecture: Charge functions for odd dimensional partitions}, we conjecture an expression of the charge function for any odd dimension. Section \ref{sec:Definition} serves as a preparation for the following proof. In section \ref{sec:Proof}, we demonstrate that the conjectured formula indeed satisfies the property of the charge function, as long as Lemma \ref{Lemma5} holds. Then in section \ref{sec:Discussion of Lemma 5}, we give a complete proof for 5D case. For higher odd dimensions, we offer a partial proof, and perform Monte Carlo sampling tests for 7D and 9D. We conclude in section \ref{sec:Summary and discussion}. The details of the proofs for Lemma \ref{Lemma1}-\ref{Lemma4} are collected in appendix \ref{sec:Lemma1-4}.  
%================================================================================================
\section{$n$-dimensional partitions}
\label{sec:n-dimensional partitions}

Let us show some basic concepts about partitions in this section following \cite{galakhov2024charging}.

$\Delta^{(n)}$ is an $n$-dimensional partition which is sometimes referred as molten crystal, if it satisfies the melting rule:
For any box $\vec{\square} \in \mathbb{Z}^n_{\geq 0}$, if there exists $\vec{\square}' \in \Delta^{(n)}$ such that $\vec{\square}' = \vec{\square} + \vec{e}_k$ for any $k = 1, 2,\cdots, n$, then $\vec{\square} \in \Delta^{(n)}$.\footnote{Our terminology follows reference~\cite{galakhov2024charging}, in which ``molten crystal" and ``partition" are precisely defined as equivalent and thus these two terms can be used interchangeably.}

We denote the canonical basis in such a positive corner of $n$-dimensional space by
\begin{equation}
\vec{e}_i = \left( \delta_{1i}, \delta_{2i}, \dots, \delta_{ni} \right), \quad i = 1, 2, \dots, n,
\end{equation}
where $\delta_{ij}$ is the Kronecker delta. We will use the notation:\\
\begin{equation}
\vec{E}\equiv\sum_{i}^{n}\vec{e}_i
\label{vecE}
\end{equation}
in our later discussion.

For an $n$-dimensional partition $\Delta^{(n)}$, the Calabi-Yau condition reads:
\begin{equation}
    \sum_{i=1}^{n}h_i=0,\label{CY condition}
\end{equation}
where the complex number $h_i$ is called weight or flavor parameter.\footnote{These parameters characterize the equivalent toric action on CY$_n$ as $(x_1,x_2,\cdots,x_n)\mapsto (e^{h_1}x_1,e^{h_2}x_2,\cdots,e^{h_n}x_n)$.} There are two sets related to the partition $\Delta^{(n)}$ that are important,
\begin{equation}
\operatorname{A}_{\Delta^{(n)}} \subset \mathbb{Z}^n_{\geq 0} \quad \text{and} \quad \operatorname{R}_{\Delta^{(n)}} \subset \mathbb{Z}^n_{\geq 0},
\end{equation}
 which respectively include all the boxes in $\mathbb{Z}^n_{\geq 0}$ that can be added to $\Delta^{(n)}$ or removed from $\Delta^{(n)}$ and the resulting partitions remain $n$-dimensional partitions. Let us introduce a projection action\footnote{In this paper, we are accustomed to denote the coordinate component of a box $\vec{\square}$ by $l_i$.}:
\begin{equation}
\label{eq:c=le}
    c:\quad \vec{\square}=\sum_{i=1}^{n}l_i \vec{e}_i\longmapsto \sum_{i=1}^{n}l_ih_i.
\end{equation}
We assume that the weight parameter $h_{1,2,\cdots,n}$ are generic complex numbers that satisfy \eqref{CY condition}. 
The projection $c$ identifies points in $\mathbb{Z}^n_{\geq 0}$ that differ by multiples of the vector $\vec{E}$. So (\ref{eq:c=le}) is not a unique decomposition of $c$ (regarded as a point in the projected space).
We can define a unique component $l'_i$ of $c$ by letting the smallest component equals to zero.\\
\begin{equation}
    l'_i\equiv l_i-min\{l_i,i={}1,2,..,n\},
\end{equation}
Then equivalently,
\begin{equation}
 c = \sum_{i=1}^{n} l'_i h_i, \quad l'_i \geq 0.
\end{equation}
We will use the notation of $\ell$ and $\ell'$ also as actions: $\ell_{i}(\vec{\square})=\ell_{i}$, $\ell_{i}'(c)=\ell_{i}'$.

%==============================================================================================
\section{Conjecture: charge functions for odd dimensional partitions}
\label{sec:Conjecture: Charge functions for odd dimensional partitions}

The charge function should 
be factored into a PRODUCT of contributions arising from all the individual boxes and all the clusters of boxes and is supposed to satisfy the following properties:
\begin{mdframed}
\begin{enumerate}
    \item \label{charge function property 1}
    $\psi_{\Delta^{(n)}}(u)$ is a meromorphic function of $u$ and only has simple poles.
    \item \label{charge function property 2}
    All the poles of $\psi_{\Delta^{(n)}}(u)$ are in one-to-one correspondence with the projected vector $c$ of the boxes $\vec{\square} \in \mathrm{A}_{\Delta^{(n)}} \cup \mathrm{R}_{\Delta^{(n)}}$.
\end{enumerate}
\end{mdframed}
After some careful consideration, we find that it is possible to construct the charge functions $\psi(u)$ for odd dimensional partitions $\Delta^{(n)}$.\footnote{The even-dimensional cases beyond 4D, which present additional nuances, have now been addressed in our subsequent work \cite{Feng:2025pft}.} Let us first review the charge functions corresponding to lower dimensional partitions following \cite{galakhov2024charging} before presenting our formula.

\vspace{2em}
\paragraph{Ordinary partitions}
For 2D partitions (Young diagrams), the Calabi-Yau condition reads
\begin{equation}
    h_1+h_2=0.
\end{equation}
The charge function for the 2D partition $\Delta^{(2)}$ is :
\begin{equation}
\psi_{\Delta^{(2)}}(u) = \frac{1}{u} 
\prod_{\vec{\square} \in \Delta^{(2)}} \varphi_1^{(2)} \left( u - c(\phi_1) \right)
\prod_{\phi_{2} \in \Delta^{(2)}} \varphi_{2}^{(2)} \left( u - c(\phi_{2}) \right)
\prod_{\phi_3 \in \Delta^{(2)}} \varphi_3^{(2)} \left( u - c(\phi_3) \right),
\end{equation}
where 
\begin{equation}
   \varphi_{1}^{(2)}(u)=\frac{1}{(u-h_1)(u-h_2)},\quad \varphi_2^{(2)}(u)=u^2,\quad \varphi_3^{(2)}(u)=\frac{1}{u^2}.
\end{equation}
and $\phi_{p}=\left \{ \vec{\square}, \vec{\square}+\vec{e}_{s_{1}}, \vec{\square}+\vec{e}_{s_{2}}, \cdots, \vec{\square}+\vec{e}_{s_{p-1}}\right \} $ is called a $p$-box cluster, and the action of $c$ on it is defined as
\begin{equation}
c(\phi_{p})=c(\vec{\square})+\sum_{i=1}^{p-1}h_{s_i}. \label{c of cluster}
\end{equation}

\vspace{2em}
\paragraph{Plane partitions}
For 3D partitions (Plane partitions), the Calabi-Yau condition is
\begin{equation}
    h_1+h_2+h_3=0.
\end{equation}
The charge function for the plane partition $\Delta^{(3)}$ is in a very important form :
\begin{equation}
\psi_{\Delta^{(3)}}(u) = \frac{1}{u} \prod_{\vec{\square} \in \Delta^{(3)}} \varphi^{(3)}_1(u - c(\vec{\square})), \label{3D charge function}
\end{equation}
where $\varphi_1^{(3)}(u)$ is the bonding factor:
\begin{equation}
\varphi^{(3)}_1(u) = \prod_{i=1}^{3} \frac{u + h_i}{u - h_i}.
\end{equation}
For 3D case, the corresponding BPS algebra is the affine Yangian of $\mathfrak{gl}_{1}$ \cite{prochazka2016}.

\vspace{2em}
\paragraph{Solid partitions}
For 4D partitions (Solid partitions), the Calabi-Yau condition reads
\begin{equation}
    h_1+h_2+h_3+h_4=0.
\end{equation}
The charge function for the solid partition $\Delta^{(4)}$ is :
\begin{equation}
\psi_{\Delta^{(4)}}(u) = \frac{1}{u} 
\prod_{\vec{\square} \in \Delta^{(4)}} \varphi_1^{(4)} \left( u - c(\phi_1) \right)
\prod_{\phi_{4} \in \Delta^{(4)}} \varphi_{4}^{(4)} \left( u - c(\phi_{4}) \right)
\prod_{\phi_5 \in \Delta^{(4)}} \varphi_5^{(4)} \left( u - c(\phi_5) \right),
\end{equation}
where 
\begin{equation}
   \varphi_{1}^{(4)}(u)=\frac{\prod_{i=1}^{4}(u+h_i)\prod_{1\le i<j\le 4}(u-h_i-h_j)}{\prod_{i=1}^{4}(u-h_i)},\quad {\varphi_4^{(4)}}(u)=\frac{1}{u^2},\quad {\varphi_5^{(4)}}(u)=u^2.
\end{equation}
and $c(\phi_p)$ still take the form as \eqref{c of cluster}.
Equivalently, 
\begin{equation}
   \varphi_{1}^{(4)}(u)=\frac{\prod_{1\le i<j<k\le 4}(u-h_i-h_j-h_k)\prod_{1\le i<j\le 4}(u-h_i-h_j)}{\prod_{i=1}^{4}(u-h_i)}.
\end{equation}

\vspace{2em}
\paragraph{Conjecture formula for arbitrary odd dimension:}
We conjecture the form of the charge function in dimension $n=2K+1,K\geq1,$

\begin{equation}
    \psi_{\Delta^{(n)}}(u)=\psi_{0}(u)\psi'_{\Delta^{(n)}}(u),
    \label{charge function general}
\end{equation}
where
\begin{equation}
    \psi_{0}=\frac{1}{u},
\end{equation}

\begin{equation}
    \psi'_{\Delta^{(n)}}(u)=\prod_{\vec{\square}\in \Delta^{(n)}}\varphi_1(u-c(\vec{\square}))\prod_{m=2}^{K}\prod_{\phi_{2m}\subset \Delta^{(n)}}\varphi_{2m}(u-c(\phi_{2m})).\label{eq:psi'}
\end{equation}
where $\phi_{2m}=\left \{ \vec{\square}, \vec{\square}+\vec{e}_{s_{1}}, \vec{\square}+\vec{e}_{s_{2}}, \cdots, \vec{\square}+\vec{e}_{s_{2m-1}}\right \} $ is a $2m$-box cluster, and $c(\phi_p)$ still follow the definition in \eqref{c of cluster}.
\begin{equation}
    c(\phi_{2m})=c(\vec{\square})+\sum_{i=1}^{2m-1}h_{s_i},
\end{equation}
and the exact form of the factors are as below:
\begin{equation}
\varphi_1(u)=\frac{\prod_{m=1}^{K}\prod_{1\leq l_1<l_2<\cdots<l_{2m}\leq 2K+1}(u-\sum_{i=1}^{2m}h_{l_{i}})}{\prod_{i=1}^{2K+1}(u-h_i)},
\end{equation}
\begin{equation}
    \varphi_{2m}(u)=\frac{1}{u}.
\end{equation}
It is easy to tell for $n=3$, i.e. $K=1$, \eqref{eq:psi'} has no contribution from clusters, and \eqref{charge function general} identifies with the known formula for plane partitions as shown in \eqref{3D charge function}.  

In the next section, we prove the formula \eqref{charge function general} proposed above indeed satisfies the properties of charge functions.
%=========================================================================================
\section{Definitions}
\label{sec:Definition}

\paragraph{Some useful sets}
We define a set of all possible partitions in $n$ dimension, and a set of points in the projected space,
\label{sec:conjecture chare function}
\begin{equation}
    P_n\coloneqq\left \{ \Delta^{(n)} \ \text{for any number of boxes} \right \} \subset 2^{\mathbb{Z}_{\geq 0}^{n} },
\end{equation}
\begin{equation}
    \mathcal{P} \coloneqq \{ c = \sum_{i=1}^{n}l_ih_i\bigm| l_i \in \mathbb{Z} \} \cong \mathbb{Z} ^{n-1}.
\end{equation}

We define the union $C_{\Delta^{(n)}}\coloneqq\operatorname{A}_{\Delta^{(n)}}\cup\  \text{R}_{\Delta^{(n)}}$ including all the possible positions to add to or remove from a partition $\Delta^{(n)}$. Since we are interested in its projection space, we define the set 
\begin{equation}
    D_{\Delta^{(n)}}\coloneqq\left \{ c=\sum_{i}l_ih_i\in\mathcal{P}\ \bigm| \ \sum_{i}l_i\vec{e}_i \in C_{\Delta^{(n)}} \right \} .
\end{equation}
Note that $\left | C_{\Delta^{(n)}} \right | =\left | D_{\Delta^{(n)}} \right | $. And we introduce another set to collect the simple pole of the partition
\begin{equation}
    \mathcal{SP} (\Delta^{(n)})\coloneqq\left \{c \ | \ c \text{ is a simple pole of } \psi_{\Delta^{(n)}}(u)\right \} .
\end{equation}

We define the set of admissible partitions at $\vec{\square}$, 
denoted by $G(\vec{\square})$, as the collection of all $n$-dimensional 
partitions for which there exists an addable or removable position that can be projected to $c(\vec{\square})$. That is,

\begin{equation}
  G(\vec{\square}) = \{\Delta^{(n)}\in P_n\ |\ \exists \,\tilde{\vec{\square}}\in C_{\Delta^{(n)}},c(\tilde{\vec{\square}})=c(\vec{\square})\}.  
\end{equation}

\vspace{1em}
\paragraph{Potential Function}
We introduce the potential function to represent the order of the pole at point $c$,
\begin{equation}
    \omega_{0,\Delta^{(n)}}(c)=\begin{cases}
    m, & \text{m-th order pole,} \\
    0, & \text{no poles or zeros,} \\
    -m, & \text{m-th order zero.}
\end{cases}
\end{equation}
The precise expression $\omega_{0,\Delta^{(n)}}(c)$ obtained from \eqref{charge function general} is as follows. The base term is:
\begin{equation}
\omega_{0,\Delta^{(n)}}(c) = \delta_{0,c} + \omega_{\Delta^{(n)}}(c).
\end{equation}
Where for a set of box \( S \subseteq \mathbb Z_{\geq0}^n \), we define:
\begin{equation}
\omega_S(c) \coloneqq \omega_{S,1}(c) + \omega_{S,\phi_{2m}}(c),
\end{equation}
with the individual components defined as follows:
\begin{equation}
\omega_{S,1}(\tilde{c}) \coloneqq \sum_{\vec{\square}\in S}\left(\sum_{i=1}^n \delta_{\tilde{c}, c(\vec{\square})+h_i} - \sum_{m=1}^K\,\sum_{0\leq l_1 \leq\cdots\leq l_{2m}\leq n} \delta_{\tilde{c},c(\vec{\square})+\sum_{i=1}^{2m}h_{l_i}}\right),
\end{equation}

\begin{equation}
\label{def:potential of cluster}\omega_{\phi_{2m}}(\tilde{c}) \coloneqq    \delta_{\tilde{c},c(\phi_{2m})},
\end{equation}

\begin{equation}
\omega_{S,\phi_{2m}}(\tilde{c}) \coloneqq \sum_{m=2}^K \,\sum_{\phi_{2m}\subseteq S} \omega_{\phi_{2m}}.
\end{equation}
We define the potential function's action on a vector $\vec{\square}$:
\begin{equation}
\omega_S(\vec{\square}) \coloneqq \omega_S(c(\vec{\square})).
\end{equation}
Note that we have translation invariance for $\omega_S(\vec{\square})$, after translating every element in set \( S \) and the  specific box position $\vec{\square}$ simultaneously along an arbitrary $n$-dimensional vector in $\mathbb{Z}^n$ to get $\tilde{S}$ and $\tilde{\vec{\square}}$ (with the requirement that all components of any element in the translated set \( \tilde{S} \) are positive), for such a translation, we have:
\begin{equation}
    \omega_{\tilde{S}}(\tilde{\vec{\square}})=\omega_S(\vec{\square}).\label{translate}
\end{equation}

For disjoint sets $S_1$ and $S_2$ (i.e., $S_1 \cap S_2 = \emptyset$), the potential function satisfies additivity with cluster correction:
\begin{equation}
\omega_{S_1 \cup S_2}(c) = \omega_{S_1}(c) + \omega_{S_2}(c) + \omega_{cluster(S_1, S_2)}(c).
\end{equation}
Among them, the term $\omega_{cluster(S_1, S_2)}$ originates from the contributions of clusters set named $\mathcal{K}_{S_1\bowtie S_2}$ that satisfy the condition specified in this section: some boxes belong to set 
$S_1$ while the other part belongs to set $S_2$.

By decomposing the potential into two contributions from single boxes and clusters, we rewrite this as:
\begin{equation}
\omega_{\text{cluster}(S_1,S_2)} = \left(\omega_{S_1\cup S_2,1} - \omega_{S_1,1} - \omega_{S_2,1}\right) + \left(\omega_{S_1 \cup S_2, \phi_{2m}} - \omega_{S_1,\phi_{2m}} - \omega_{S_2,\phi_{2m}}\right)\label{eq:omega cluster}.
\end{equation}
The first term vanishes identically, while the second term can be expressed as a sum over clusters crossing $S_1$ and $S_2$, defined as set of relevant clusters $\mathcal{K}_{S_1\bowtie S_2}$ :
\begin{equation}
\omega_{S_1 \cup S_2, \phi_{2m}} - \omega_{S_1,\phi_{2m}} - \omega_{S_2,\phi_{2m}} = \sum_{\phi_{2m}\in \mathcal{K}_{S_1\bowtie S_2}}\omega_{\phi_{2m}}.
\end{equation}
Substituting this back to (\ref{eq:omega cluster}), we obtain the final expression of $\omega_{{cluster}}$:
\begin{equation}
\omega_{{cluster}(S_1,S_2)}(\vec{\square}) = \sum_{\phi_{2m}\in \mathcal{K}_{S_1\bowtie S_2}}\omega_{\phi_{2m}}(\vec{\square}).
\end{equation}

\vspace{1em}
\paragraph{$\textbf{$d$}$-neighbor} 
A configuration $c$ is called a $d$-neighbor of $c'$ (where $d < n$), denoted as $c' \xrightarrow{d} c$ (we sometimes omit $d$ and write $c' \leftrightarrow c$), if and only if:
\begin{equation}
c = c' + \sum_{i=1}^d h_{n_i},
\end{equation}
for some index set $\{n_i\}$ with $n_i \in \{1,2,\dots,n\}$. It is straightforward to show that the neighborhood relation is symmetric in the sense of dimension complement:
\begin{equation}
c' \xrightarrow{d} c \iff c \xrightarrow{n-d} c'.
\end{equation}

Note that we have properties:
\begin{enumerate}
    \item Given $\vec{\square} $:
    \begin{equation}
    \exists S, \, \omega_S(c') \neq \omega_{S-\vec{\square}}(c') \implies c(\vec{\square}) \leftrightarrow c'\label{eq:neighbor pr1},
    \end{equation}
    
    \item The neighborhood relation is equivalent to the following conditions:
    \begin{equation}
    c \leftrightarrow c' \iff \forall i, \, l_i'(c - c') \in \{0,1\}.
    \end{equation}
    which is further equivalent to:
    \begin{equation}
    \forall i,j, \, l_i(c - c') - l_j(c - c') \in \{-1,0,1\}\label{eq:neighbor pr2}.
    \end{equation}
\end{enumerate}

\vspace{1em}
\paragraph{Bisect operation} 
Now we define an operation that divides the partition set $\Delta^{(n)}$ into two subsets, $L{(\Delta^{(n)},\vec{\square})}$ and $\Delta^{(n)}-L{(\Delta^{(n)},\vec{\square})}$. 

\begin{figure}[!htbp]
    \centering
    % 子图 (a): Bisect Operation
    \begin{subfigure}[b]{0.32\linewidth}
        \centering
        \includegraphics[width=\linewidth]{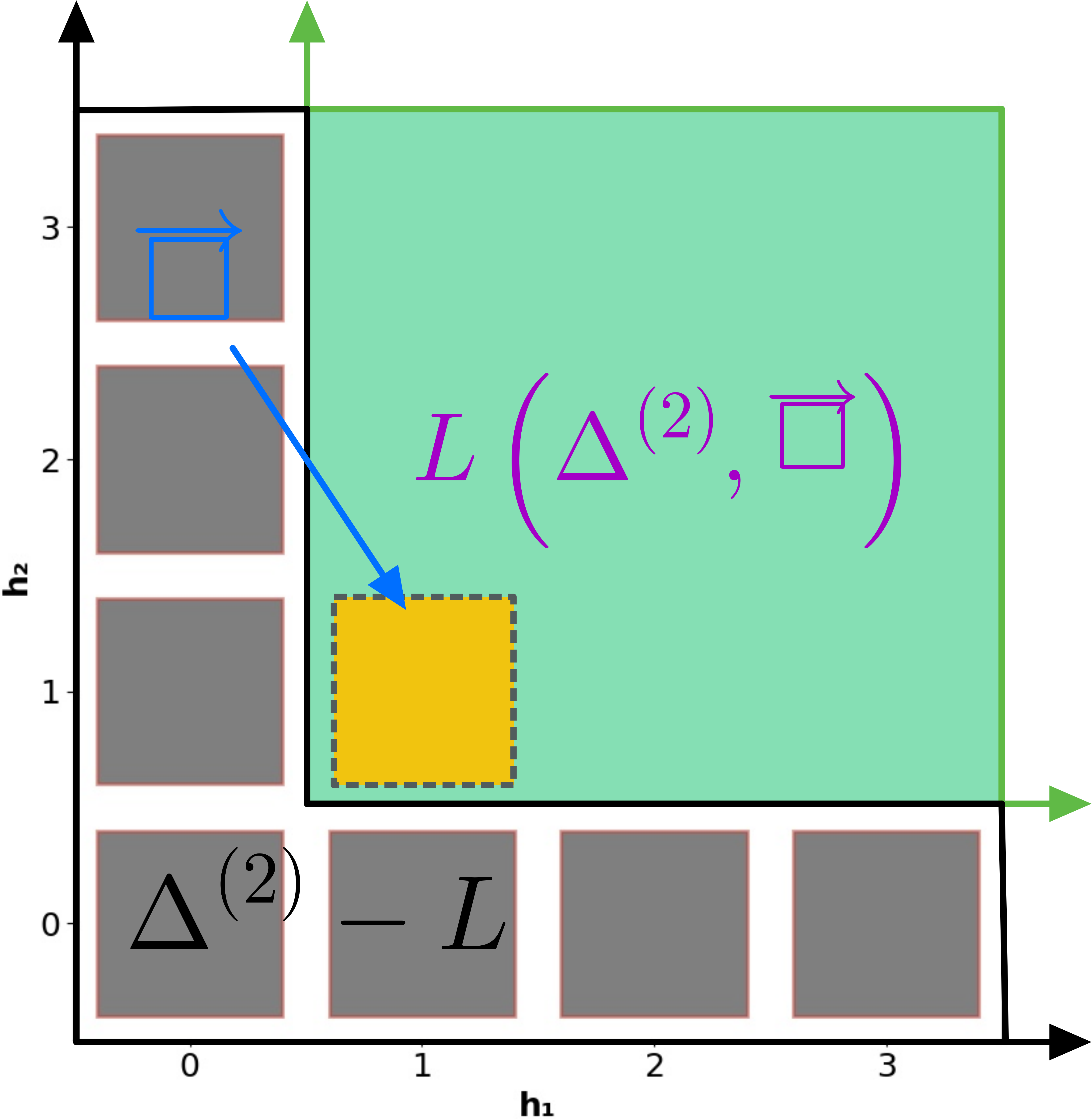} 
        \caption{Bisect operation $L$.}
        \label{fig:bisect}
    \end{subfigure}
    \hfill 
    % 子图 (b): Hypercube (新加的)
    \begin{subfigure}[b]{0.32\linewidth}
        \centering
        \includegraphics[width=\linewidth]{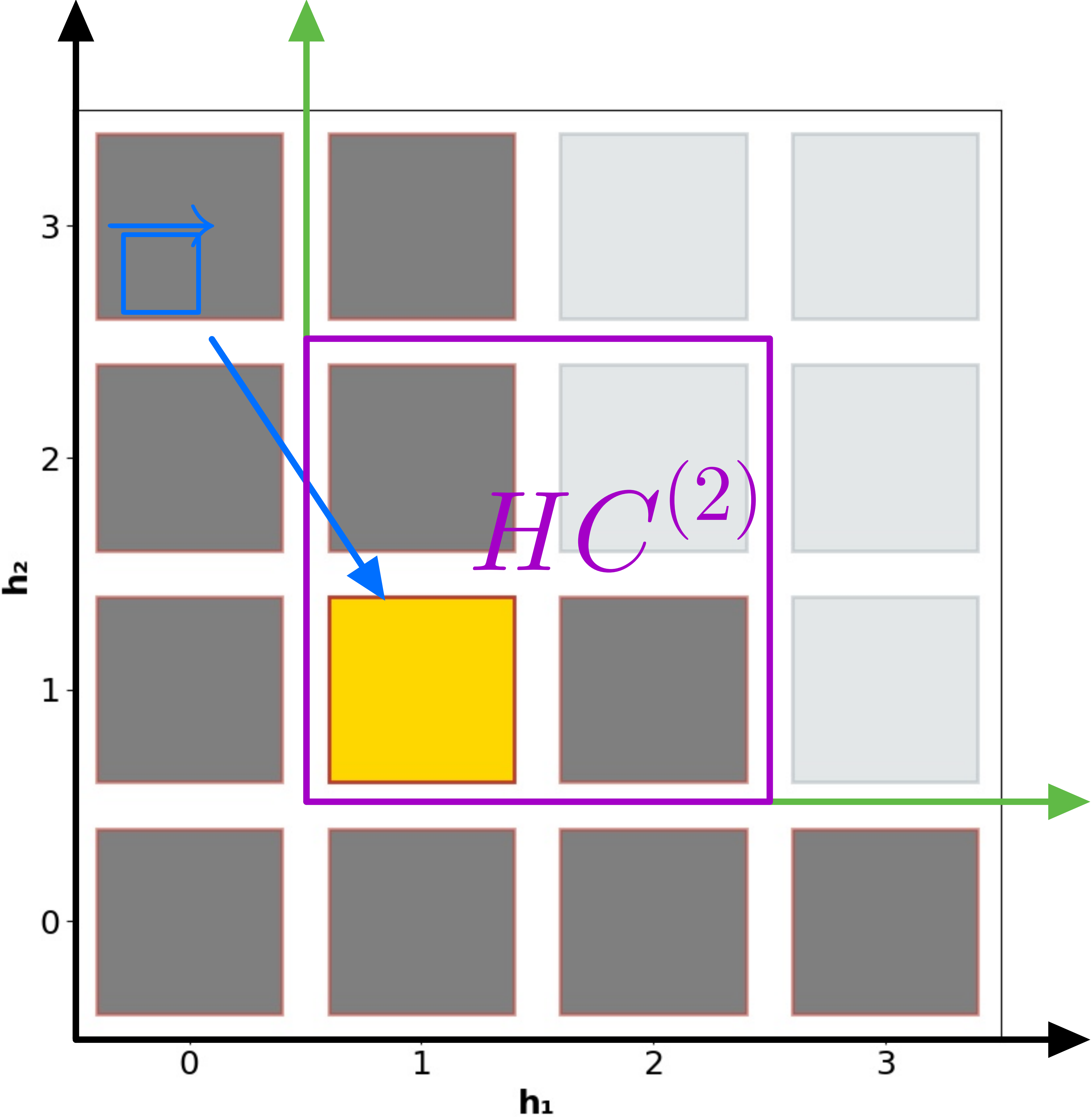} % 请确保文件名正确
        \caption{Hypercube $HC^{(2)}$.} 
        \label{fig:hypercube} 
    \end{subfigure}
    \hfill
    % 子图 (c): Surface Set
    \begin{subfigure}[b]{0.32\linewidth}
        \centering
        \includegraphics[width=\linewidth]{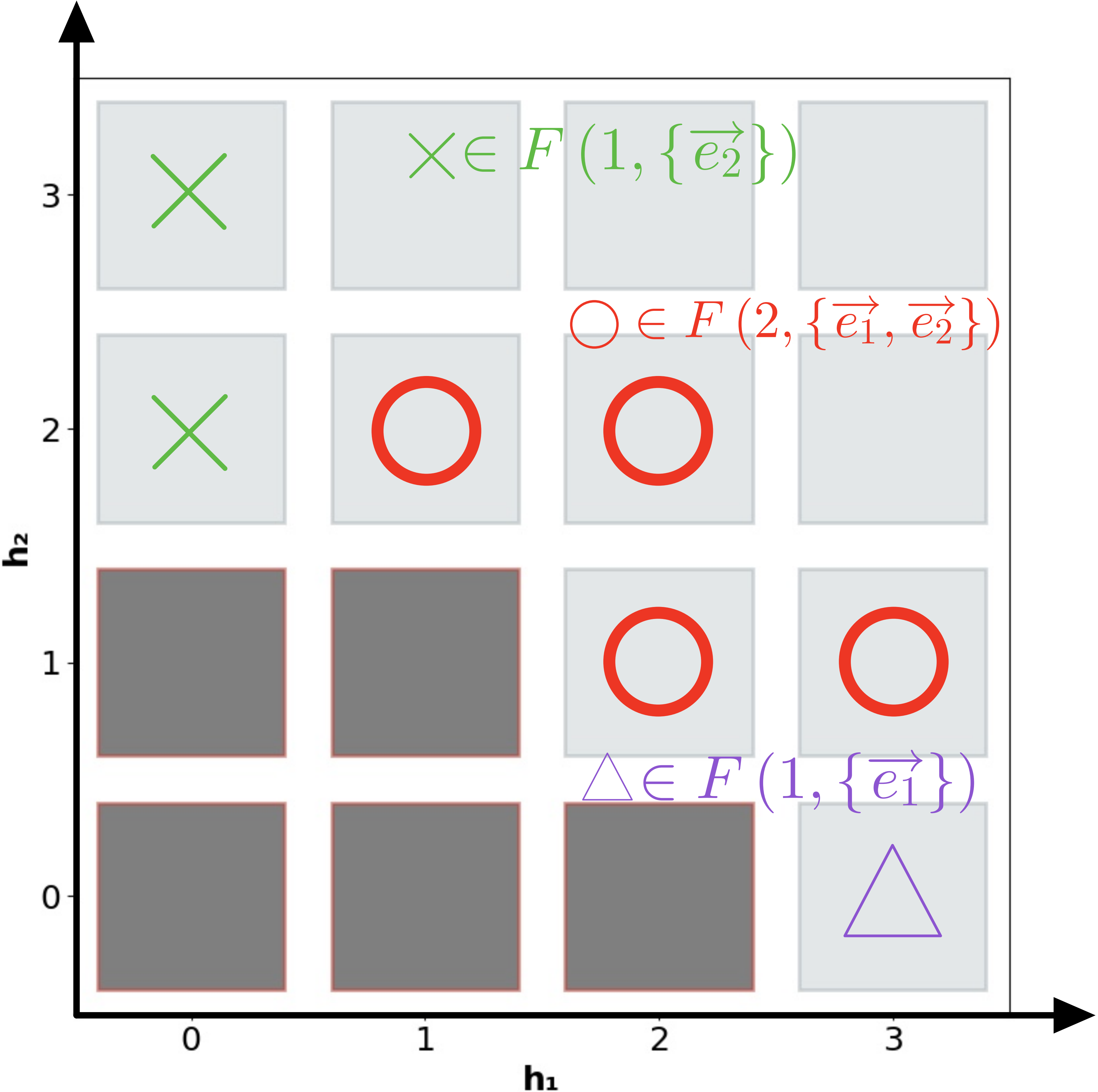} 
        \caption{Surface sets $F$.} 
        \label{fig:surface} 
    \end{subfigure}
    
    \caption{{Schematic diagrams of definitions in 2D. (a) illustrates the bisect operation $L(\Delta^{(2)}, \vec{\square})$ (green region) and the remaining part (grey region). (b) shows a hypercube $HC^{(2)}$ (purple frame) with its origin (yellow box). (c) depicts different surface sets $F(d, \{\vec{e}_{n_i}\})$ marked by different symbols.}}
    \label{fig:define} 
\end{figure}

For a partition $\Delta^{(n)}$, define $L$ as operation that yields a collection of boxes $L$, where each box satisfies the condition that all its components are greater than or equal to the corresponding components of box $\vec{\square}=(l_1,l_2,...,l_n)$ :
\begin{equation}
\label{def:bisect}L{(\Delta^{(n)},\vec{\square})}\coloneqq \Delta^{(n)}\cap \{\tilde{\vec{\square}}=\sum_{i=1}^n\tilde{l_i}\vec{{e_i}}\,|\,\tilde{l_i}-l_i\geq0,\,\forall\,i\}. \end{equation}
As illustrated in Fig.~\ref{fig:bisect}.

\vspace{1em}
\paragraph{Hypercube}
A $d$-dimensional hypercube $HC$ with its origin $\vec{\square}$ is defined as a subspace with $2^d$  elements in $\mathbb{Z}_{\geq0}^n$ (see Fig.~\ref{fig:hypercube} for a 2D example):
\begin{equation}
\label{def:HC}HC^{(d)}(\vec{\square},\{e_{n_i}\}_{i=1}^d)\coloneqq\{\tilde{\vec{\square}}\,|\,\tilde{\vec{\square}}=\vec{\square}+\sum_{i=1}^d \delta_i \vec{e}_{n_i},\delta_i=0,1\}.
\end{equation}
\vspace{1em}
\paragraph{Surface set}
\label{def:surface}
We now define the set of positions located on the surface (which means it is the position of the smallest components among those unoccupied position with the same $c$) of the partition. For a fixed dimension $d\leq n$ and a given set of basis vectors $\{e_{n_i}\}$ (where $i=1,\dots,d$, $n_i \in \{1,2,\dots,n\}$), we define the set of \emph{$d$-dimensional surface points} as
\begin{equation}
F(d,\{\vec{e}_{n_i}\}) \coloneqq \left\{ 
  \vec{\square} \,\middle|\, 
  \begin{aligned}
    &\vec{\square} = \sum_{i=1}^n l_i \vec{e}_i \notin \Delta^{(n)}; \\
    &\,l_{n_j} \neq 0 \ (\forall j\in\{1,\dots,d\}),\ l_k = 0 \ (\forall k\notin{\{n_j\}}), (d<n), \\
    &\vec{\square} - \vec{E} \in \Delta^{(n)},\quad (d=n),
  \end{aligned}
\right\}
\end{equation}
where $\vec{E}\equiv\sum_{i}^{n}\vec{e}_i$ as defined in \eqref{vecE}. That is to say, for $\vec{\square}\in F(d,\{\vec{e}_{n_i}\})$, we mean $\vec{\square}$ is a position on the surface of partition with $n-d$ components equal to zero and the set of nonzero directions is $\{\vec{e}_{n_i}\}$.

Taking the union over all possible dimensions and all index sets, given $\Delta^{(n)}$ we define the complete set of surface positions:
\begin{equation}
F \coloneqq \bigcup_{d=0}^{n} \bigcup_{\{n_i\}} F(d,\{\vec{e}_{n_i}\}).
\end{equation}
3 different types of surface sets in 2D example are shown in Fig.~\ref{fig:surface}. Note that there exists a natural bijection between the boxes (positions) in $F$ and the elements of the projection space $\mathcal{P}$; thus, the set $F$ precisely captures the surface structure of interest.\\

\vspace{1em}
\paragraph{Conjecture}
Using the notation introduced above, we can rewrite our conjecture as:
\begin{mdframed}
    \begin{enumerate}
        \item $\psi_{\Delta^{(n)}}(u)$ has only simple poles.
        \item $D_{\Delta^{(n)}} = \mathcal{SP}_{\Delta^{(n)}}$, which is equivalent to
              $\Delta^{(n)} \in G(\vec{\square}) \, \,\Longleftrightarrow \, \,
                  \omega_{0,\Delta^{(n)}}(\vec{\square}) = 1.$ 
    \end{enumerate}
    \end{mdframed}
%==========================================================================================
\section{Proof of the Conjecture}
\label{sec:Proof}
In this section, we start with introducing some lemmas, and prove the conjecture basing on these lemmas.
\subsection{Lemmas}
\begin{Lemma}
\label{Lemma1}
(proved in Appendix \ref{subsec:Lemma1}) For $\forall$ $\Delta^{(n)}$ and  $\vec{\square}$, after bisect operation $L$, the remaining boxes in $\Delta^{(n)}-L$ still form a partition.
\begin{equation}\Delta^{(n)}-L(\Delta^{(n)},\vec{\square})\in P_n\,.
\end{equation}
\end{Lemma}

\begin{Lemma}
\label{Lemma2}
(proved in Appendix \ref{subsec:Lemma2}) For $\forall$ $\Delta^{(n)}$ and $\vec{\square}\in F(d,\{\vec{e}_{n_i}\})$, The potential at $\vec{\square}$ of original partition $\Delta^{(n)}$ is equal to that in processed partition ${\Delta^{(n)} - \tilde{L}+\widetilde{HC}}$.

\begin{equation}
\omega_{\Delta^{(n)}} (\vec{\square}) = \omega_{\Delta^{(n)} - \tilde{L}+\widetilde{HC} }(\vec{\square}).
\end{equation}
Where for later convenience, we denote.\\
\begin{equation}
\widetilde{HC}(d,\Delta^{(n)},
\vec{\square},\{\vec{e}_{n_i}\}_{i=1}^{d}) \equiv HC^{(d)} \left(\vec{\square} - \sum_{i=1}^{d} \vec{e}_{n_i} , \{\vec{e}_{n_i}\}\right) \cap \Delta^{(n)} , \vec{\square}\in F(d,\{\vec{e}_{n_i}\}),\end{equation}
\begin{equation}
\widetilde{L}(\Delta^{(n)},
\vec{\square}) \equiv L (\Delta^{(n)}, \vec{\square} - \sum_{i=1}^{d} \vec{e}_{n_i}).
\end{equation}
To simplify the notation and improve readability in the subsequent discussions, we will omit the explicit arguments in parentheses for the sets and operators defined above when the context is unambiguous. Specifically, we will denote the hypercube $\widetilde{HC}(d, \Delta^{(n)}, \vec{\square}, \{\vec{e}_{n_i}\}_{i=1}^d)$ simply as $\widetilde{HC}$, and the bisected partition component $\tilde{L}(\Delta^{(n)}, \vec{\square})$ simply as $\tilde{L}$.
In the definition here, we specify an 
$n$-dimensional vector $\vec{\square}$ as the position for performing the bisect operation $\tilde{L}$ and generating the hypercube $\widetilde{HC}$.
\end{Lemma}

\begin{Lemma}
\label{Lemma3}
(proved in Appendix \ref{subsec:Lemma3}) For $\forall$ $\Delta^{(n)}$ and $\,\vec{\square}\in F(d,\{\vec{e}_{n_i}\})$.
The potential contributed by the clusters in $\tilde{\mathcal{K}} \equiv \mathcal{K}_{\widetilde{HC} \bowtie (\Delta^{(n)}-\tilde{L})}$ is zero.
\begin{equation}
\omega_{cluster(\widetilde{HC},\Delta^{(n)}-\tilde{L})}(\vec{\square})=0\label{eq:intersect}.
\end{equation}
\end{Lemma}

\begin{Lemma}
\label{Lemma4}
(proved in Appendix \ref{subsec:Lemma4}) For $\forall$ $\Delta^{(n)}$ and $\vec{\square}\in F(d,\{\vec{e}_{n_i}\})$, $0<d<n$.
\begin{equation}
\omega_{0,\Delta^{(n)}-\tilde{L}} (\vec{\square}) = 0
\end{equation}
\end{Lemma}

\begin{Lemma}
(proved for $n=5$ and discussed for higher n in section \ref{sec:Discussion of Lemma 5}) For $\forall$ $\Delta^{(n)}\subseteq HC^{(d)}(\vec{0},\{\vec{e}_{n_i}\})$.

\begin{empheq}[
  left={\omega_{0,\Delta^{(n)}}\left(\sum_{i=1}^d\vec{e}_{n_i}\right) = \empheqlbrace}  % 左侧共享大括号
]{align}
1,  &\quad \text{if } \Delta^{(n)} \in G\left(\sum_{i=1}^d\vec{e}_{n_i}\right),
\label{5(a)} \tag{5a} \\  
\le 0,  &\quad \text{if } \Delta^{(n)} \notin G\left(\sum_{i=1}^d\vec{e}_{n_i}\right).
\label{5(b)}  \tag{5b}  
\end{empheq}
\label{Lemma5}
\end{Lemma}

\subsection{Main proof}
\noindent \textbf{Outline of the proof strategy.} Before detailing the inductive step, we briefly outline the logical flow. The proof relies on mathematical induction on the total number of boxes, $|\Delta^{(n)}|$. To determine the pole structure at a specific target position $\vec{\square}$, we utilize the \textit{bisect operation} and the \textit{hypercube} construction defined previously. The strategy involves decomposing the potential function of the full partition into two independent parts using \textbf{Lemma 2}: a "background" partition term ($\Delta^{(n)} - \tilde{L}$) and a local hypercube term ($\widetilde{HC}$). \textbf{Lemma 3} ensures that the cross-terms (cluster interactions) between these two parts vanish, effectively decoupling the problem. We then show via \textbf{Lemma 4} that the background partition contributes trivially to the potential, leaving the local hypercube as the sole determinant. Finally, \textbf{Lemma 5} provides the explicit evaluation of the hypercube's potential, confirming that it correctly reproduces the required pole order (1 for admissible positions, $\leq 0$ otherwise).

\begin{proof}We proceed by mathematical induction. The conjecture is immediate in $|\Delta^{(n)}| = 0$. Assume that our conjecture holds for all cases where $|\Delta^{(n)}| < N$. Now, consider the case where $|\Delta^{(n)}| = N$. Given a vector $\vec{\square} \in F(d,\{\vec{e}_{n_i}\})$, $\vec{\square}=\sum_{i=1}^{n}l_i\vec{e_i}$, we analyze the potential function $\omega_{\Delta^{(n)}}(\vec{\square})$ in two scenarios based on the dimension $d$.

We begin by writing the potential function with Lemma \ref{Lemma2}:
\begin{equation}
\omega_{\Delta^{(n)}}(\vec{\square}) = \omega_{\Delta^{(n)} - \tilde{L}+\widetilde{HC}}(\vec{\square})=\omega_{\Delta^{(n)} - \tilde{L}}(\vec{\square})+\omega_{\widetilde{HC}}(\vec{\square})+\omega_{cluster(\widetilde{HC},\Delta^{(n)}-\tilde{L})}(\vec{\square}) .
\end{equation}
The third term is zero according to Lemma \ref{Lemma3}, then
\begin{equation}
\begin{aligned}
\omega_{0,\Delta^{(n)}}(\vec{\square}) 
&= \delta_{0,c(\vec{\square})} + \omega_{\Delta^{(n)}}(\vec{\square}) \\
&= \delta_{0,c(\vec{\square})} + \omega_{\Delta^{(n)} - \tilde{L}}(\vec{\square}) + \omega_{\widetilde{HC}}(\vec{\square}) \\
&= \omega_{0,\Delta^{(n)} - \tilde{L}}(\vec{\square}) + \omega_{\widetilde{HC}}(\vec{\square}).
\label{eq:main proof,1}
\end{aligned}
\end{equation}
First, consider the case where $d=n$, noted that:
\begin{equation}
\tilde{\vec{\square}}\equiv\vec{\square}-\vec{E}\in A_{\Delta^{(n)}-\tilde{L}},
\end{equation}
since $\tilde{\vec{\square}} + \vec{e}_i \in \tilde{L}$, $\tilde{\vec{\square}} - \vec{e}_i \in \Delta^{(n)}-\tilde{L}$ and $\tilde{\vec{\square}}\notin \Delta^{(n)}-\tilde{L}$. We also have $|\Delta^{(n)}-\tilde{L}|<N$ because at least $\tilde{\vec{\square}}$ is removed. And we know from Lemma \ref{Lemma1} that $\Delta^{(n)}-\tilde{L}$ is a partition. Our conjecture at lower level guarantees that $\omega_{0,\Delta^{(n)} - \tilde{L}}(\vec{\square})=\omega_{0,\Delta^{(n)} - \tilde{L}}(\tilde{\vec{\square}})=1$. (\ref{eq:main proof,1}) is now:
\begin{equation}
\omega_{0,\Delta^{(n)}}(\vec{\square}) = 1 + \omega_{\widetilde{HC}}(\vec{\square}) = 1 + \omega_{\widetilde{HC}-[\vec{\square}-\vec{E}]}(\vec{E})=\omega_{0,\widetilde{HC}-[\vec{\square}-\vec{E}]}(\vec{E})\label{eq:main proof,2}.
\end{equation}
The second equal holds because translate invariance see (\ref{translate}), $\widetilde{HC}-[\vec{\square}-\vec{E}]$ means that the new hypercube obtained by translating the hypercube along $\vec{\square}-\vec{E}$ and its origin exactly located at $\vec{0}$.

Second, consider the case where $d<n$. The first term in (\ref{eq:main proof,1}) equals zero By Lemma \ref{Lemma4} :
\begin{equation}
\omega_{0,\Delta^{(n)}}(\vec{\square}) =  \omega_{\widetilde{HC}}(\vec{\square})= \omega_{\widetilde{HC}-[\vec{\square}-\sum_{i=1}^d\vec{e}_{n_i}]}(\sum_{i=1}^d\vec{e}_{n_i})=\omega_{0,\widetilde{HC}-[\vec{\square}-\sum_{i=1}^d\vec{e}_{n_i}]}(\sum_{i=1}^d\vec{e}_{n_i}).
\end{equation}
Which has the same form as (\ref{eq:main proof,2})
Finally, by Lemma \ref{Lemma5}, \begin{equation}
    \omega_{0,\widetilde{HC}-[\vec{\square}-\sum_{i=1}^d\vec{e}_{n_i}]}(\sum_{i=1}^d\vec{e}_{n_i})=1\iff\widetilde{HC}-[\vec{\square}-\sum_{i=1}^d\vec{e}_{n_i}]\in G(\sum_{i=1}^d\vec{e}_{n_i}).
\end{equation}

On the other hand, for any $\vec{\square} \in F(d, \{\vec{e}_i\})$,  $\Delta^{(n)}\in G(\vec{\square})\iff\widetilde{HC}-[\vec{\square}+\sum_{i=1}^d\vec{e}_{n_i}]\in G(\sum_{i=1}^d\vec{e}_{n_i})$ because the melting rule only involves those boxes in $\widetilde{HC}$. As a result, we have the equivalence $\omega_{\Delta^{(n)}}(\vec{\square}) = 1 \iff \Delta^{(n)} \in G(\vec{\square})$. Since we can see that $\omega_{\Delta^{(n)}}(\vec{\square}) \le 1$ from Lemma \ref{Lemma5}, this proves the conjecture.
\end{proof}
%=============================================================================================
\section{Discussion of Lemma 5}
\label{sec:Discussion of Lemma 5}
For convenience, we repeat the statement of Lemma \ref{Lemma5}:

\noindent\textbf{Lemma 5} 

$\forall$ $\Delta^{(n)}\subseteq HC^{(d)}(\vec{0},\{\vec{e}_{n_i}\})$,

\begin{equation}
\omega_{0,\Delta^{(n)}}(\sum_{i=1}^d\vec{e}_{n_i})=
\begin{cases}
1 & \Delta^{(n)}\in G(\sum_{i=1}^d\vec{e}_{n_i}), \\
\le 0 & \Delta^{(n)}\notin G(\sum_{i=1}^d\vec{e}_{n_i}).
\end{cases}
\end{equation}

\begin{figure}[p]
    \centering
    \includegraphics[width=0.85\textwidth]{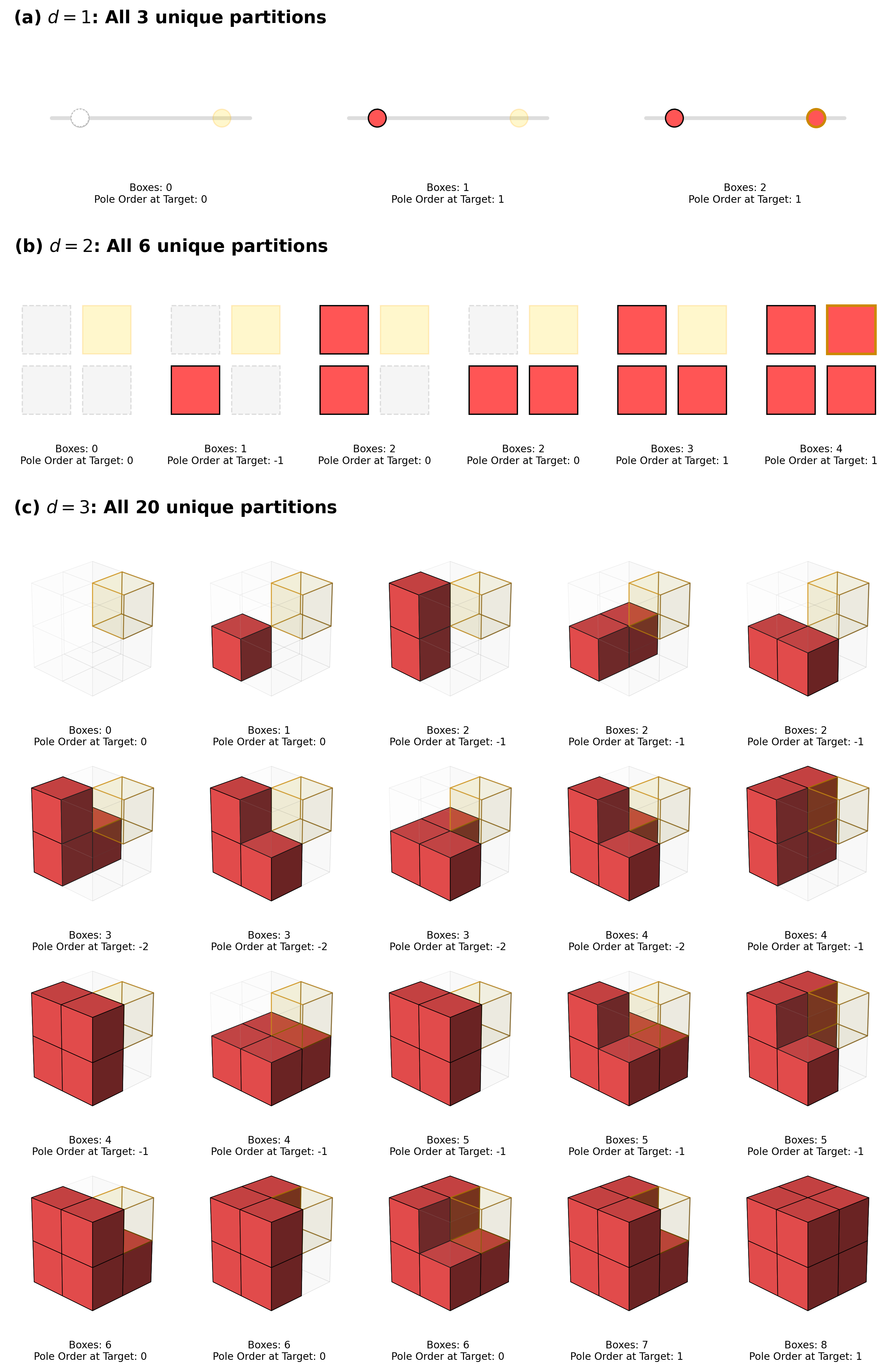}
    \caption{
Exhaustive visualizations of all unique partitions within low-dimensional hypercubes. 
        (a) The 3 unique partitions for the $d=1$ hypercube.
        (b) The 6 unique partitions for the $d=2$ hypercube.
        (c) The 20 unique partitions for the $d=3$ hypercube.
            For each case, red cubes represent occupied boxes, while the target position is highlighted with a semi-transparent orange color. The number of boxes and the calculated pole order at the target position are specified below each configuration.}
    \label{fig:partitions_d123}
\end{figure}

To study Lemma \ref{Lemma5}, we write $HC^{(d)}\left(\vec{0},\{\vec{e}_{n_i}\}_{i=1}^d \right )=HC^{(d)}$ for short. We first claim that for a partition $\Delta^{(n)}\subseteq HC^{(d)}$, the condition $\Delta^{(n)}\in G(\sum_{i=1}^d\vec{e}_{n_i})$ is met only by a specific set of partitions. For the case where the dimension of the hypercube equals the dimension of the space ($d=n$), these partitions are the empty set, the single-box partition at the origin, the fully occupied hypercube, and the hypercube missing only the corner-most box $\vec{E}$. For $d<n$, they are the fully occupied hypercube and the one missing the corner-most box. This can be verified by the melting rule.

As a visual confirmation, Fig.~\ref{fig:partitions_d123} provides an exhaustive enumeration of all unique partitions for hypercubes of dimensions $d=1, 2,$ and $3$. Across all these cases, it is evident that the pole order at the target position is exactly 1 only for the partitions corresponding to the fully occupied hypercube and the hypercube with one box missing at the target position. This observation is in perfect agreement with our claim and the properties outlined in Lemma 5.

\subsection{Analytical verification of Lemma 5}
In this subsection, we provide an analytical proof for Lemma 5a, specifically focusing on the pole cancellation mechanism within the hypercube structure. Our goal is to verify that the total potential function $\omega_{0, \Delta^{(n)}}$ evaluates to one when the partition is an admissible hypercube configuration.

We begin with the base cases. It is straightforward to verify from the definition that:
\begin{equation}
    \omega_{0,\{\vec{0}\}}(\vec{E}) = \omega_{0,\emptyset}(\vec{E}) = 1,
\end{equation}
and for the full hypercube minus the target boxes:
\begin{equation}
    \omega_{0,HC^{(d)}}(\sum_{i=1}^d\vec{e}_{n_i}) = \omega_{0,HC^{(d)}-\{\sum_{i=1}^d \vec{e}_{n_i}\}}(\sum_{i=1}^d\vec{e}_{n_i}).
\end{equation}
Our task is to prove that $\omega_{0,HC^{(d)}} = 1$. The total pole order at the target position is the sum of contributions from all relevant clusters within the hypercube. Due to the highly symmetric geometry of the hypercube $HC^{(d)}$, the number of neighbors at a specific distance (or co-dimension) is determined simply by combinatorial factors.

Specifically, the number of $(n-m)$-neighbors of the target vector $c=\sum_{i=1}^d h_{n_i}$ within the hypercube is given by the binomial coefficient:
\begin{equation}
    \left| \{c(\vec{\square}) \mid c \xrightarrow{m} \sum_{i=1}^d h_{n_i}, \vec{\square} \in HC^{(d)} \} \right| = \binom{d}{m}.
\end{equation}
According to the structure of the charge function, each $(n-m)$-neighbor (where $1 \le m \le n-1$) contributes a factor of $(-1)^{m+1}$ to the pole order either. Therefore, calculating the potential function reduces to evaluating the alternating sum of these binomial coefficients.

We distinguish two cases based on the relationship between the partition dimension $n$ and the hypercube dimension $d$:

\paragraph{Case 1: $n \neq d$.}
The summation covers all $m$ from 1 to $d$. Using the binomial expansion of $(1-1)^d$, we have:
\begin{equation}
    \omega_{HC^{(d)}}(\sum_{i=1}^d h_{n_i}) = \sum_{m=1}^{d} \binom{d}{m} (-1)^{m+1} = 1.
\end{equation}
Thus, $\omega_{0, HC^{(d)}} = \omega_{HC^{(d)}} = 1$.

\paragraph{Case 2: $n = d$.}
In this case, the summation range is restricted to $m \in [1, n-1]$ because the term $m=n$ corresponds to the origin, which is treated separately in the base potential. The sum becomes:
\begin{equation}
    \omega_{HC^{(n)}}(\vec{E}) = \sum_{m=1}^{n-1} \binom{n}{m} (-1)^{m+1}=0.
\end{equation}
Finally, incorporating the base term $\omega_{0, \emptyset} = 1$, we obtain the total potential:
\begin{equation}
    \omega_{0,HC^{(n)}}(\vec{E}) = 1 + \omega_{HC^{(n)}}(\vec{E}) = 1.
\end{equation}
This completes the analytical proof of Lemma 5a, demonstrating that the combinatorial structure of the hypercube ensures the correct pole order through the properties of binomial coefficients.

\subsection{Numerical proof of Lemma 5}
For the d-dimensional hypercubes satisfying the requirements in Lemma \ref{Lemma5}, we numerically enumerated all unique partitions for $d$=1 to 5, proving that Lemma \ref{Lemma5} holds rigorously for all 5D case (totally $3+6+20+168+7581$ unique cases). 

For the 7 and 9 dimensional case, Because the Dedekind number M(7) \& M(9) is too large ($2.4147\times10^{12}$ and a 42-digit value calculated in 2023 \cite{D9}), we performed Monte Carlo sampling, verifying partitions with different numbers of boxes for $d$=1 to $n$. For all existing sampling results, the upper bound predictions of Lemma \ref{Lemma5} for potential fully meet the requirements.

\begin{figure}[!h]
    \centering
        \begin{subfigure}[b]{0.49\textwidth}
        \centering
        \includegraphics[width=\textwidth]{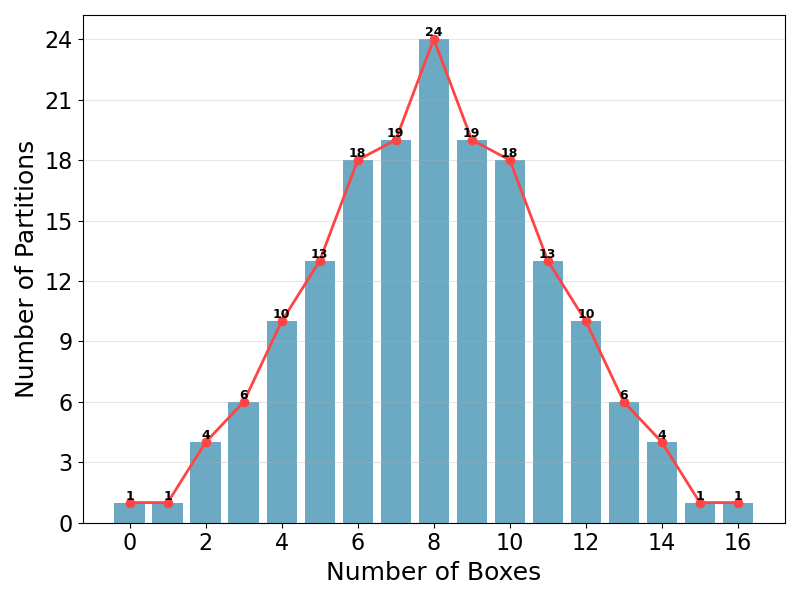}
        \caption{${HC}^{(4)}$ partition distribution}
        \label{subfig:4d_dist}
    \end{subfigure}
        \begin{subfigure}[b]{0.49\textwidth}
        \centering
        \includegraphics[width=\textwidth]{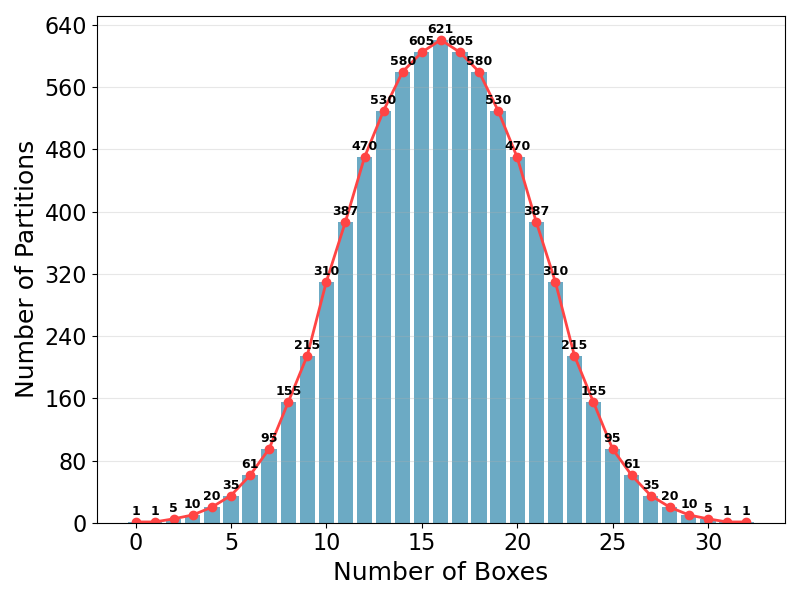}
        \caption{${HC}^{(5)}$ partition distribution}
        \label{subfig:5d_dist}
    \end{subfigure}
    \begin{subfigure}[b]{0.49\textwidth}
        \centering
        \includegraphics[width=\textwidth]{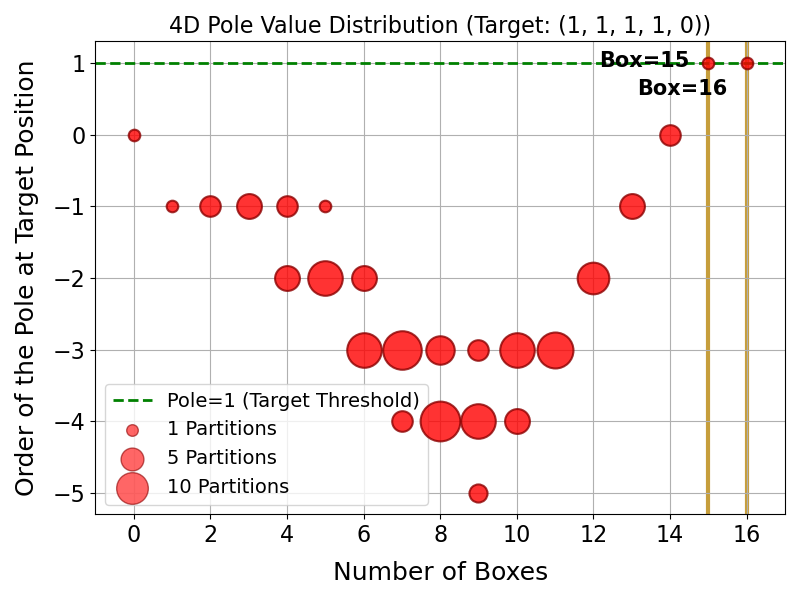}
        \caption{${HC}^{(4)}$  pole order at target position}
        \label{subfig:4d_bubble}
    \end{subfigure}
    \hfill
    \begin{subfigure}[b]{0.49\textwidth}
        \centering
        \includegraphics[width=\textwidth]{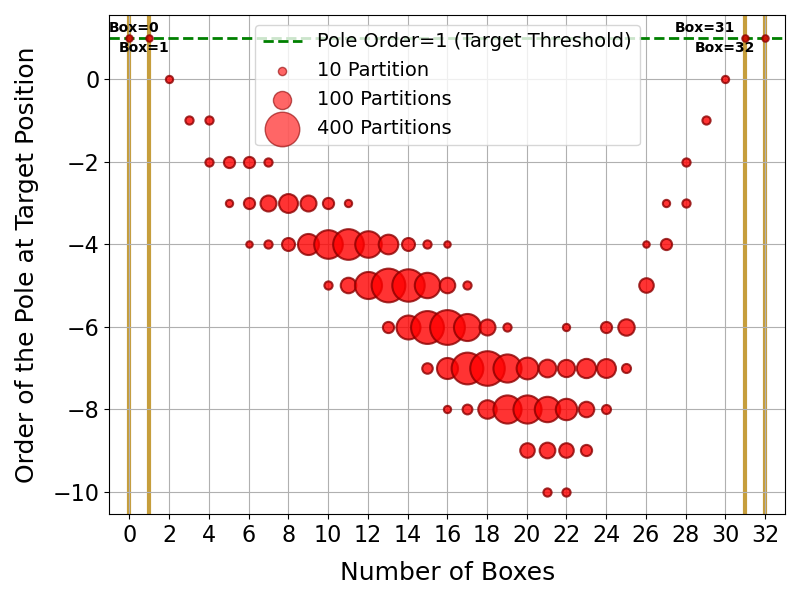}
        \caption{${HC}^{(5)}$  pole order at target position}
        \label{subfig:5d_bubble}
    \end{subfigure}
    
    \caption{Numerical results for $n=5$. Upper panel: Distributions of unique partition counts vs. box numbers for ${HC}^{(4)}$ and ${HC}^{(5)}$, respectively. 
Lower panel: Target pole orders for 
${HC}^{(4)}$ and ${HC}^{(5)}$ partitions with different box numbers. Bubble size denotes the count of unique partitions for each (box number, pole order) pair. For $d=4$, pole order=1 only for 15/16 boxes (fully occupied/one missing box); for 
$d=5$, pole order=1 for 0/1/31/32 boxes.}
    \label{fig:four_grid}
\end{figure}

Fig.\ref{subfig:4d_dist}\&\ref{subfig:5d_dist} shows the distribution of the number of unique partitions as a function of the number of boxes composing the partitions, for ${HC}^{(4)}$ and ${HC}^{(5)}$ respectively. The maximum number of unique partitions occurs when the number of boxes is $2^{d-1}$.

Fig.\ref{subfig:4d_bubble}\&\ref{subfig:5d_bubble} show the order of the pole at the target position corresponding to partitions with different numbers of constituent boxes in ${HC}^{(4)}$ and ${HC}^{(5)}$. The size of the bubble represents the quantity of unique partitions with particular number of boxes and target pole order. It can be observed that for $d=4$, the pole order is 1 only for 15 and 16 boxes case (the fully occupied case and the case with one missing box), while for $d=5$, the pole order is 1 when the number of boxes is 0/1/31/32 in $d=5$ case.

For higher dimensions such as 7D and 9D, the total number of unique partitions is computationally intractable to enumerate exhaustively. To verify our formula in these cases, we therefore employ a Monte Carlo sampling method to generate a statistical ensemble of random, valid partitions. Our randomization process is a direct sequential growth algorithm, which we describe here. The procedure to generate a single valid partition $\Delta$ of a target size $N$ is as follows:
\begin{enumerate}
    \item \textbf{Initialization:} The process begins with an empty partition, $\Delta_0 = \emptyset$.
    \item \textbf{Iteration:} At each step $k$ (from $0$ to $N-1$), we first identify the set of all ``addable" boxes, $A_{\Delta_k}$. A box is considered addable if it is not currently in the partition $\Delta_k$ and its addition would result in a new valid partition, $\Delta_{k+1}$, that still satisfies the melting rule.
    \item \textbf{Random Selection:} A single box is chosen uniformly at random from this set of addable boxes $A_{\Delta_k}$.
    \item \textbf{Growth:} The selected box is added to the partition, forming $\Delta_{k+1} = \Delta_k \cup \{\text{selected box}\}$.
    \item \textbf{Termination:} This iterative process is repeated until the partition reaches the desired size, $|\Delta_N| = N$.
\end{enumerate}
This entire generation procedure is repeated many times for each target box count $N$ to create a large ensemble of random partitions. The properties of the charge function, such as the average contribution of its terms, are then computed over this ensemble. This method provides an effective way to explore the vast configuration space of higher-dimensional partitions where direct enumeration is not feasible.

\begin{figure}[!htbp]
    \centering
    \begin{subfigure}[b]{0.49\linewidth}
        \centering
        \includegraphics[width=\linewidth]{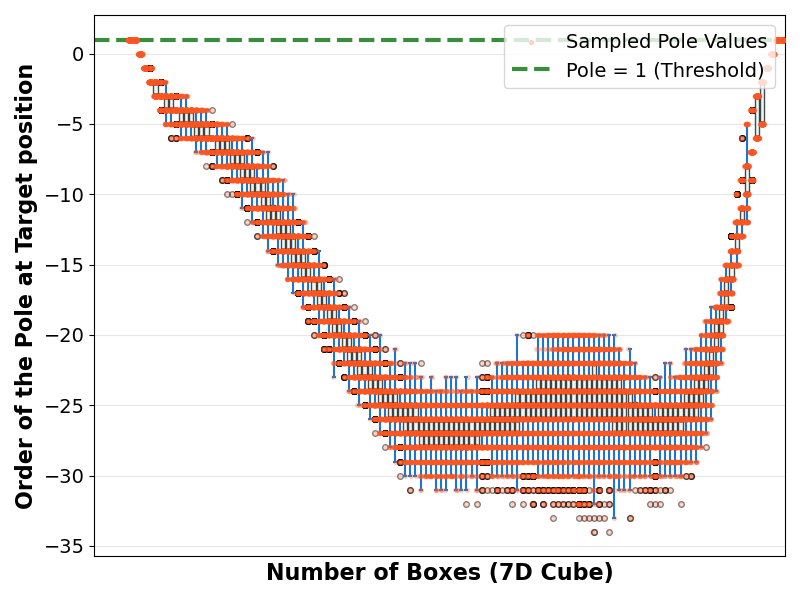} 
        \centering
        \caption{Monte Carlo sampling results for \\
        \centering$d=n=7$}
        \label{fig:7d}
    \end{subfigure}
    \hfill 
    \begin{subfigure}[b]{0.49\linewidth}
        \includegraphics[width=\linewidth]{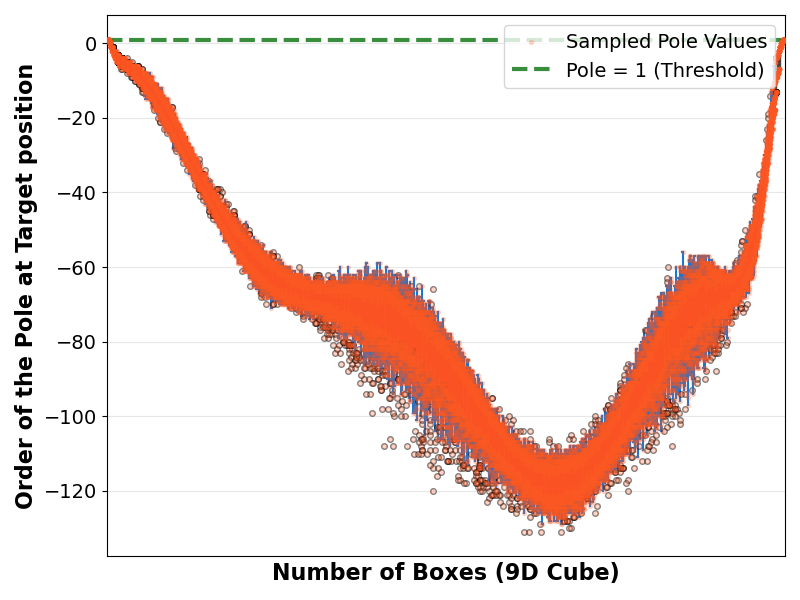} 
        \caption{Monte Carlo sampling results for\\ 
        \centering$d=n=9$} 
        \label{fig:9d} 
    \end{subfigure}
    \caption{Monte Carlo sampling results for 7D and 9D case, where the sample results are in good agreement with the description of Lemma 5.}
    \label{fig:7d9d} 
\end{figure}

Fig.\ref{fig:7d}\&\ref{fig:9d} shows the Monte Carlo sampling results for $n=d=7$ and $n=d=9$ case, it can be clearly seen that the results for the samples are in good agreement with the description of Lemma \ref{Lemma5}, providing numerical confidence for our proof.

\begin{figure}[htbp]
    \centering
    \begin{subfigure}[b]{0.32\textwidth}
        \includegraphics[width=\textwidth]{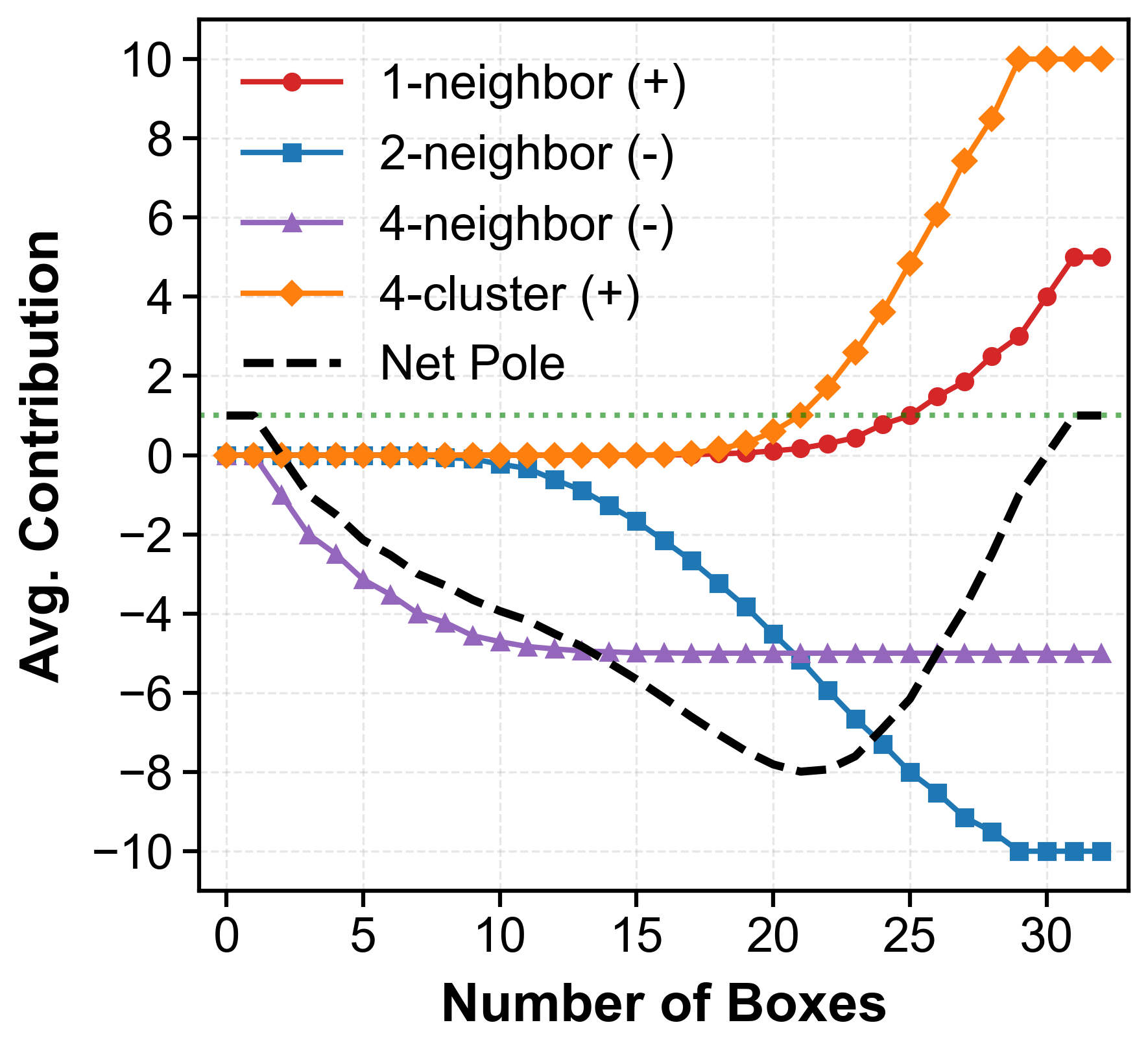}
        \caption{$n=5$ Case}
        \label{fig:contrib_5d}
    \end{subfigure}
    \hfill % 在子图之间创建水平间距
    \begin{subfigure}[b]{0.32\textwidth}
        \includegraphics[width=\textwidth]{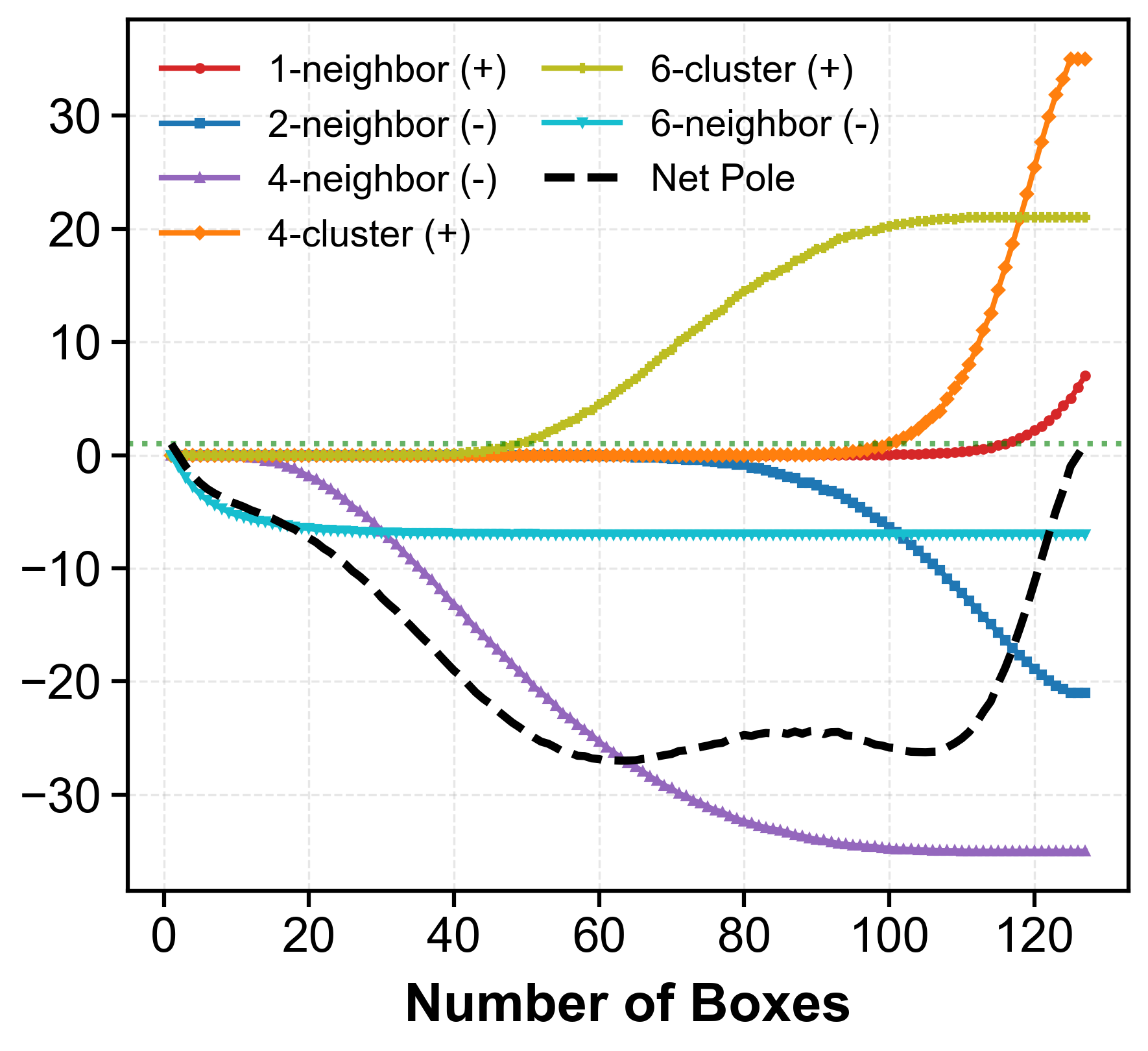}
        \caption{$n=7$ Case}
        \label{fig:contrib_7d}
    \end{subfigure}
    \hfill % 在子图之间创建水平间距
    \begin{subfigure}[b]{0.32\textwidth}
        \includegraphics[width=\textwidth]{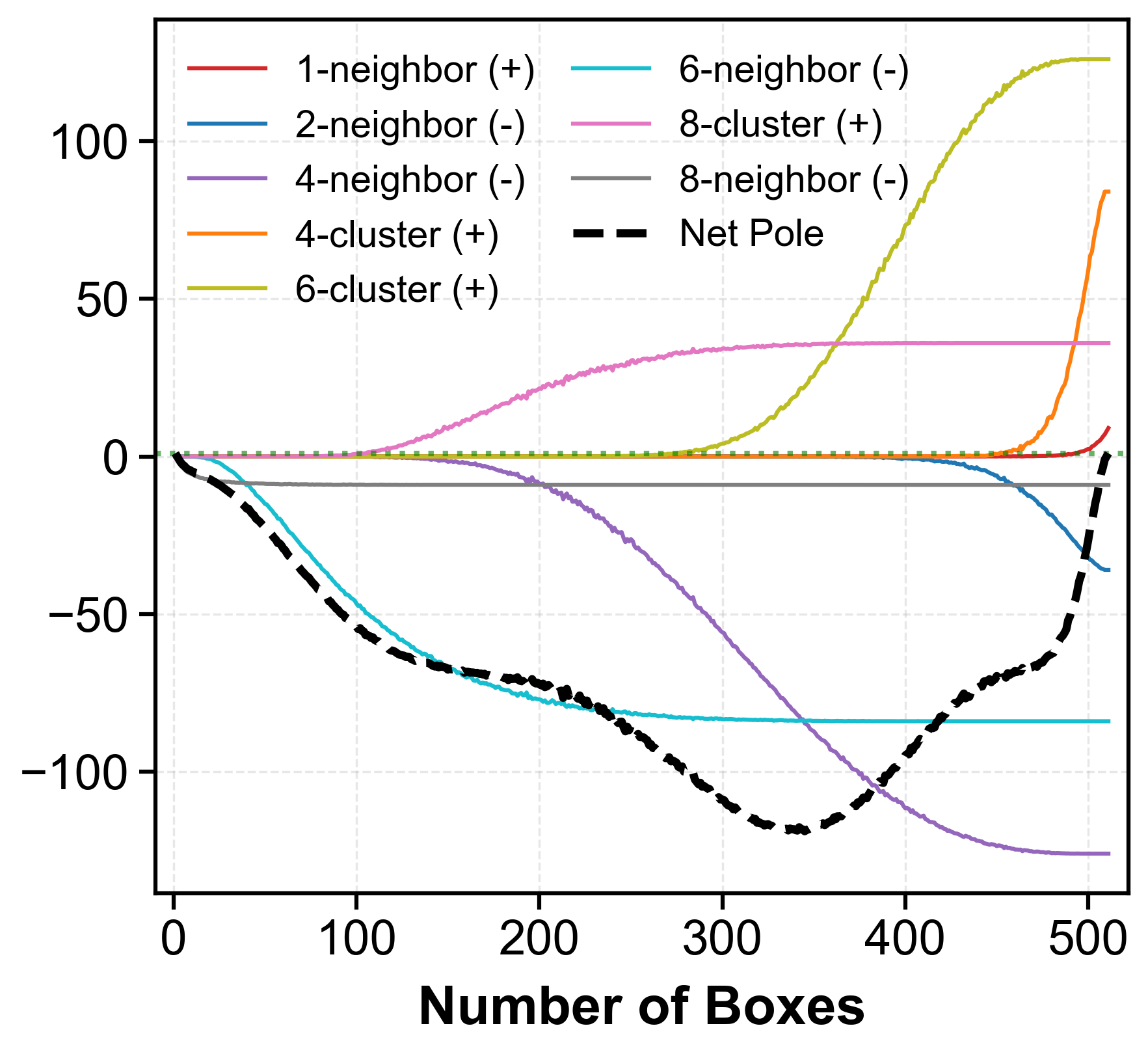}
        \caption{$n=9$ Case}
        \label{fig:contrib_9d}
    \end{subfigure}
    \caption{
        Average contribution of different terms in the charge function formula to the target pole order, plotted against the number of boxes in the partition for 5D, 7D, and 9D hypercubes. For each dimension, the values are averaged over all possible unique partitions (5D) or a large number of Monte Carlo samples (7D and 9D). The black dashed line represents the net pole order, which is the sum of all contributions plus the base pole of 1. A key observation is the role of higher-order cluster terms (e.g., 4-cluster in 5D, 6-cluster in 7D, and 8-cluster in 9D), which become significant only for partitions with a large number of boxes, providing the necessary positive contribution to ensure the net pole order remains consistent with our conjecture.
    }
    \label{fig:term_contributions}
\end{figure}

To further investigate the validity of our proposed charge function for higher odd dimensions, we analyze the contribution of each term in our formula. Since for a fixed number of boxes there can be many different partition configurations, we compute the average contribution of each term to the target pole order. The results for 5D, 7D, and 9D are presented in Fig.~\ref{fig:term_contributions}.

In the 5D case (Fig.~\ref{fig:contrib_5d}), where we performed an exhaustive enumeration, a clear cancellation mechanism is visible. For a small number of boxes, the positive contribution from the 1-neighbor term is primarily balanced by the negative 2-neighbor term. As the number of boxes increases, the 4-neighbor term provides a significant negative contribution, which is precisely counteracted by the emergence of the 4-cluster term. This delicate balance ensures the net pole order (black dashed line) behaves as predicted.

For the 7D (Fig.~\ref{fig:contrib_7d}) and 9D (Fig.~\ref{fig:contrib_9d}) cases, which are based on Monte Carlo sampling, we observe a similar pattern. The higher-order cluster terms (e.g., the 5-cluster and 7-cluster terms) only become substantial when the partition is nearly full. Their positive contributions are crucial for canceling the large negative values from higher-order neighbor terms (e.g., 6-neighbor and 8-neighbor terms). This numerical evidence strongly supports that these higher-order cluster terms are not only present but are essential for the consistency of our conjectured formula in higher dimensions.

\begin{figure}
    \centering
    \includegraphics[width=0.63\linewidth]{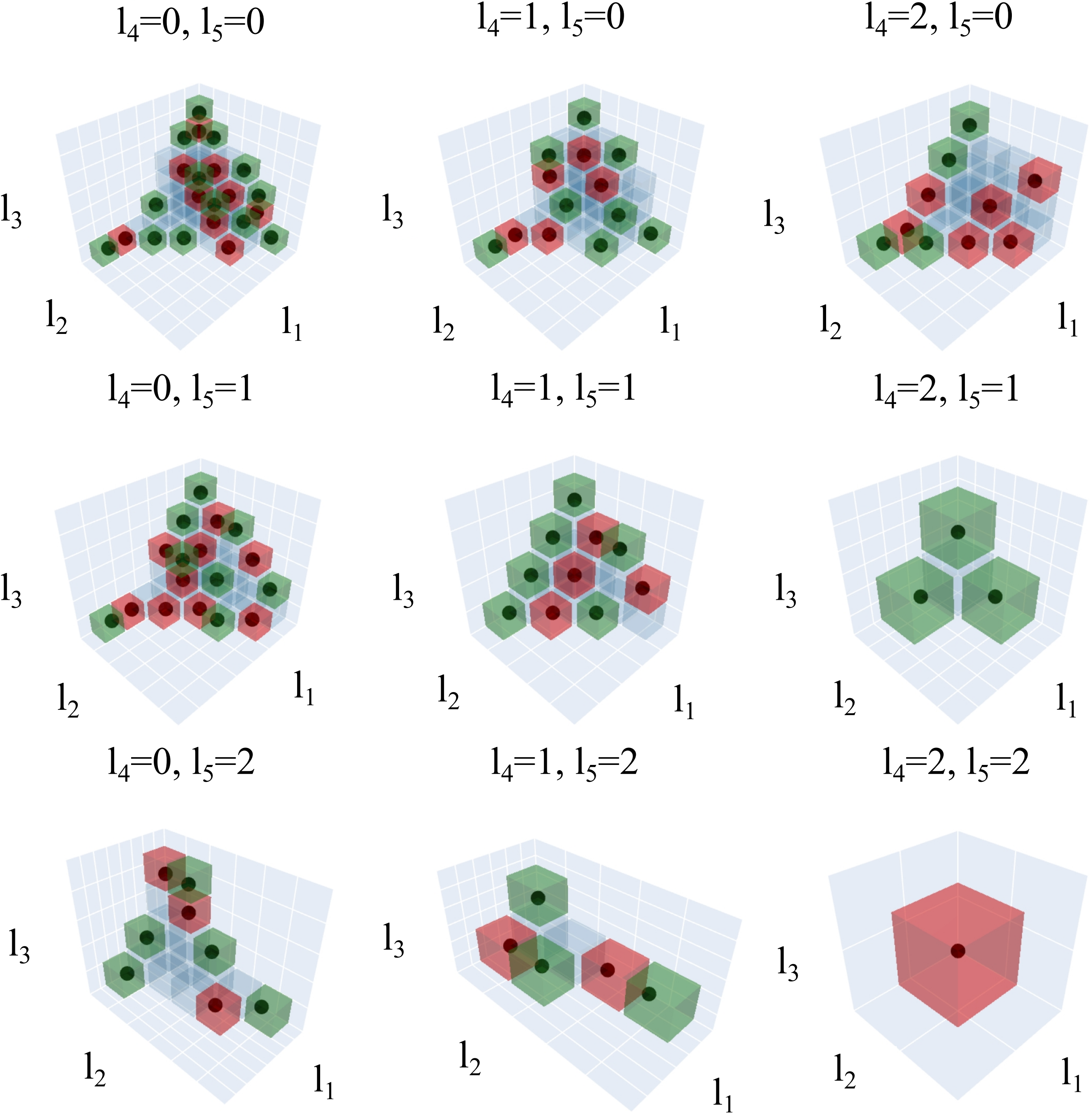}
    \caption{Visualization of a special $n=5$ partition (200 boxes). Each subplot shows its projection onto $\vec{e}_1,\vec{e}_2,\vec{e}_3$, with horizontal rightward as increasing $\vec{e}_4$ and vertical downward as increasing $\vec{e}_5$. Red/green squares denote the positions which can add new boxes or remove existing boxes (melting rule), and black dots represent simple poles of the charge function. Their perfect coincidence confirms consistent judgments between the two methods, verifying the universality of our method for general cases.}
    \label{fig:compare}
\end{figure}

Fig.\ref{fig:compare} presents the visualization of a special $n=5$ partition (comprising 200 boxes). Each subplot shows its projection onto the first three dimensions ($h_1,h_2,h_3$), where the horizontal rightward direction represents the increasing order of $h_4$, and the vertical downward direction represents the increasing order of $h_5$. In the figure, the red and green squares denote the positions which can add new boxes or remove existing boxes via the melting rule, while the black dots represent the simple poles of the charge function. It can be observed that the two perfectly coincide, indicating consistent judgment results between the two methods and verifying the universality of our method for general cases.

\section{Summary and discussion}
\label{sec:Summary and discussion}

In this paper, we achieve a breakthrough by successfully constructing the charge function \eqref{charge function general} that is universally applicable to any odd-dimensional partitions. A critical foundation of our proof lies in Lemma \ref{Lemma5}, whose validity is indispensable for ensuring the rigor and correctness of the entire theoretical framework. Only when Lemma \ref{Lemma5} holds can the charge function effectively fulfill its designed role. To consolidate this foundational result, we not only prove for Lemma \ref{5(a)}, but also conduct comprehensive, numerical validations to corroborate the reliability of Lemma \ref{5(a)} and Lemma \ref{5(b)}.

We perform an exhaustive enumeration of all unique partitions in $d$-dimensional hypercubes for dimensions $d = 1$ to $5$ in $\mathbb{Z}_{\geq0}^5$. This exhaustive search confirms that Lemma \ref{Lemma5} holds for all 5D cases (encompassing a total of 7778 unique cases), thus finishing the proof for 5D case. For higher-dimensional cases, exhaustive enumeration becomes computationally intractable due to the exponential growth of the partition space ($2^{2^n}$). Instead, we adopt a Monte Carlo sampling approach, which systematically verifies partitions with varying numbers of boxes across the dimensional range from $d = 1$ to the target dimension $n$ (i.e., $7$ and $9$), ensuring broad coverage of possible partitions.

Notably, all numerical results, whether from exhaustive enumeration ($n=5$) or Monte Carlo sampling ($n = 7, 9$), consistently demonstrate that the upper bound predictions for the potential derived from Lemma \ref{Lemma5} fully meet the required theoretical conditions. Building on this validated foundation of Lemma \ref{Lemma5}, we further rigorously prove that the constructed charge function satisfies the crucial properties \ref{charge function property 1} and \ref{charge function property 2}, which are essential for accurately capturing the correct pole structure of the system. Collectively, our theoretical construction of the charge function, Lemma \ref{5(a)}, and extensive numerical verifications (covering $d = 1$ to $5$ via exhaustive enumeration and $d = 7, 9$ via Monte Carlo sampling) provide compelling evidence for the validity and robustness of our proposed framework.

However, it is not easy to generalize our result (\ref{charge function general}) to even dimensional cases for the following two reasons. 

\noindent\textbf{1. Odd-order product terms induce asymmetric distribution}

\eqref{eq:psi'} can be formally written in an approximation form,\\
\begin{equation}
    \psi'_{\Delta^{(n)}}(u)\sim \prod_{\vec{\square}\in\Delta^{(n)}}\varphi(u-c(\vec{\square}))\label{eq:psi' simple},
\end{equation}
where
\begin{equation}
\varphi(u)=\frac{\prod_{m=1}^{K}\prod_{1\leq l_1<l_2<\cdots<l_{2m}\leq 2K+1}(u-\sum_{i=1}^{2m}h_{l_{i}})}{\prod_{m=1}^{K}\prod_{1\leq l_1<l_2<\cdots<l_{2m-1}\leq 2K+1}(u-\sum_{i=1}^{2m-1}h_{l_{i}})}.
\end{equation}

For a large partition, (\ref{eq:psi' simple}) differs from \eqref{eq:psi'} only in the contribution from clusters at the surface of the partition. There are odd number of integers $s \in [1,n-1]$ if n is even, which implies that terms of the form 
\begin{equation}
    \prod\limits_{1\leq l_1<l_2<\cdots<l_{s}\leq n}\left(u-\sum_{i=1}^{s}h_{l_{i}}\right) ,
\end{equation}

cannot be evenly distributed between the numerator and denominator in \eqref{eq:psi' simple}.

\noindent\textbf{2. Pole contribution breaks sign symmetry for even $n$}

The pole contribution from $\vec{\square}$ must be $+1$ to a 1-neighbor of $c(\vec{\square})$, and $-1$ to an $(n-1)$-neighbor. For even $n$, this implies the pole contribution from $\vec{\square}$ to a $d$-neighbor of $c(\vec{\square})$ cannot follow the conjectured pattern $(-1)^d$.\\

\vspace{8pt}

In spite of the even dimensional issue mentioned above, which we leave for future work,
our result serves as a foundational step toward constructing BPS algebras for Calabi-Yau $n$-folds. Just as the charge function for $n=3$ leads to the affine Yangian of $\mathfrak{gl}_1$ and the $n=4$ case motivates the Solid Algebra, our formula provides the necessary eigenvalue data to bootstrap the algebra generators for $n=5$ and beyond.

\vspace{8pt}

$\\$
\noindent {\it Acknowledgements.} 
The authors thank Yutaka Matsuo, Jean-Emile  Bourgine,  Rui-Dong Zhu,  Zifan Chen and Martijn Kool
for helpful discussions.  K.Z. (Hong Zhang) is supported by a classified fund of Shanghai city.
%===================================================================================
\appendix
\section{Proofs of Lemma 1-4}
\label{sec:Lemma1-4}
\subsection{Lemma 1}
\label{subsec:Lemma1}
For $\forall$ $\Delta^{(n)}$ and $\vec{\square}$, after bisect operation $L$, the remaining boxes in $\Delta^{(n)}-L$ still form a partition.
\begin{equation}\Delta^{(n)}-L(\Delta^{(n)},\vec{\square})\in P_n\,.
\end{equation}
\begin{proof}
 For $\forall\ \tilde{\vec{\square}} \in \Delta^{(n)} - L(\Delta^{(n)},\vec{\square})$ and $i\in \{1,2,...,n\}$, the melting rule in $\Delta^{(n)}$ implies:\\
\begin{equation}
    \tilde{\vec{\square}}-\vec{e}_i\in \Delta^{(n)},
\end{equation}
We also have:\\
\begin{equation}
    \tilde{\vec{\square}}-\vec{e}_i\notin L(\Delta^{(n)}, \vec{\square}),
\end{equation}
 due to the definition of $L$ (\ref{def:bisect}). Thus we have\\
 \begin{equation}
   \tilde{\vec{\square}}-\vec{e}_i\in\Delta^{(n)} - L(\Delta^{(n)}, \vec{\square}) , 
 \end{equation}
which is equivalent to the melting rule of $\Delta^{(n)} - L(\Delta^{(n)},\vec{\square})$, thus proving the Lemma by definition.
\end{proof}
\subsection{Lemma 2}
\label{subsec:Lemma2}

 For $\forall$ $\Delta^{(n)}$ and $\vec{\square}\in F(d,\{\vec{e}_{n_i}\})$, The potential at $\vec{\square}$ of original partition $\Delta^{(n)}$ is equal to that in processed partition ${\Delta^{(n)} - \tilde{L}+\widetilde{HC}}$.
\begin{equation}
\omega_{\Delta^{(n)}} (\vec{\square}) = \omega_{\Delta^{(n)} - \tilde{L}+\widetilde{HC} }(\vec{\square}).
\end{equation}

\begin{proof}
First, we start with the inclusion relation that the hypercube \(\widetilde{HC}\) is a subset of \(\widetilde{L}\), i.e.,
\begin{equation} \widetilde{HC} \subset \widetilde{L}. \end{equation}
Based on this inclusion, we can decompose \(\widetilde{L}\) into the disjoint union of \(\widetilde{L} - \widetilde{HC}\) and \(\widetilde{HC}\), which gives
\begin{equation} \widetilde{L} = (\widetilde{L} - \widetilde{HC}) \cup \widetilde{HC}. \end{equation}

Next, consider an arbitrary vector \(\tilde{\vec{\square}} \in \widetilde{L} - \widetilde{HC}\). For all indices \(i\), the component-wise condition holds due to $\tilde{\vec{\square}}
\in \tilde{L}$:
\begin{equation} \widetilde{l}_i - (l_i - 1) \ge 0. \end{equation}
Since \(\vec{\square} \notin \widetilde{HC}\), there exists at least one index \(i\) such that the component difference satisfies
\begin{equation} \widetilde{l}_i - (l_i - 1) \ge 2 \Leftrightarrow \widetilde{l}_i - l_i \ge 1. \end{equation}
On the other hand, because \(\vec{\square} \notin \Delta^{(n)}\) and using the melting rule, there exists some index \(j\) where
\begin{equation} \widetilde{l}_j - l_j \le -1. \end{equation}
Combining these two results, we find that there exist indices \(i, j\) such that the difference of component differences is bounded below by 2:
\begin{equation} (\widetilde{l}_i - l_i) - (\widetilde{l}_j - l_j) \ge 2. \end{equation}
(\ref{eq:neighbor pr2}) then implies that the $n-1$ vector \(c(\vec{\square})\) is not a neighbor of \(c(\tilde{\vec{\square}})\), denoted as
\begin{equation} c(\vec{\square}) \nleftrightarrow c(\tilde{\vec{\square}}). \end{equation}

Now, take any intermediate set \(I\) satisfying \(\Delta^{(n)} - \widetilde{L} + \widetilde{HC} \subsetneq I \subset \Delta^{(n)}\). By the neighborhood non-equivalence established above and (\ref{eq:neighbor pr1}), the potential function \(w_I(\vec{\square})\) remains unchanged when removing any \(\tilde{\vec{\square}} \in \widetilde{L} - \widetilde{HC}\), i.e.,
\begin{equation} w_I(\vec{\square}) = w_{I - \tilde{\vec{\square}}}(\vec{\square}) \quad \forall \ \tilde{\vec{\square}} \in \widetilde{L} - \widetilde{HC}. \end{equation}

By iteratively removing all elements of \(\widetilde{L} - \widetilde{HC}\) from \(I\) and using the invariance of the potential function, we finally obtain the desired equality:
\begin{equation} w_{\Delta^{(n)}}(\vec{\square}) = w_{\Delta^{(n)} - \widetilde{L} + \widetilde{HC}}(\vec{\square}). \end{equation}
\end{proof}

\subsection{Lemma 3}
\label{subsec:Lemma3}

 For $\forall$ $\Delta^{(n)}$ and $\vec{\square}\in F(d,\{\vec{e}_{n_i}\})$.
The potential contribute by the clusters in $\tilde{\mathcal{K}}=\mathcal{K}_{\widetilde{HC} \bowtie (\Delta^{(n)}-\tilde{L})}$ is zero,
\begin{equation}
\omega_{cluster(\widetilde{HC},\Delta^{(n)}-\tilde{L})}(\vec{\square})=0.
\end{equation}

\begin{proof}
Suppose $\phi_{2m}$ is a cluster contributing to
\begin{equation}
\omega_{\text{cluster}(\widetilde{HC}, \Delta^{(n)} - \tilde{L})} (\vec{\square}).
\end{equation}
We first prove the following key claim:
\begin{equation}
\vec{\square}_c \in \Delta^{(n)} - \tilde{L} \quad \text{and} \quad \exists! k, \, \vec{\square}_c + \vec{e}_k \in \tilde{L}.
\label{eq:Lemma3,result1}
\end{equation}

We prove (\ref{eq:Lemma3,result1}) by contradiction. Suppose for contradiction that the claim fails. If $\vec{\square}_c \in \widetilde{HC}\subset\tilde{L}$, it is straightforward to show that $\{\vec{\square}_c + \vec{e}_i\} \in \tilde{L}$ for $\forall i$ by the definition of $L$ (\ref{def:bisect}). This implies $\phi_{2m} \cap (\Delta^{(n)} - \tilde{L}) = \emptyset$, which contradicts the assumption that $\phi_{2m}$ contributes to the cluster potential (as clusters require non-trivial intersection with both sets). 

Thus, we must have $\vec{\square}_c \notin \widetilde{HC}$, which in turn implies $\exists k$ such that $\vec{\square}_c + \vec{e}_k \in \tilde{L}$ ( $\phi_{2m} \cap (\Delta^{(n)} - \tilde{L}) \neq \emptyset$). For $\forall j \neq k$, note that $l_k(\vec{\square}_c + \vec{e}_j) = l_k(\vec{\square}_c) < l_k(\vec{\square})$, and by the characterization of $\tilde{L}$, this gives $\vec{\square}_c + \vec{e}_j \notin \tilde{L}$. Combining these two results, we conclude $\exists! k$ such that $\vec{\square}_c + \vec{e}_k \in \tilde{L}$, and since $\vec{\square}_c \notin \widetilde{HC}$, we also have $\vec{\square}_c \in \Delta - L$. This completes the proof of (\ref{eq:Lemma3,result1}).

From above, we immediately derive the component-wise relation for the vector difference:
\begin{equation}
l_i (\vec{\square} - \vec{\square}_c) =
\begin{cases}
1 & i=k, \\
\le 0 & i \neq k,
\end{cases}
\end{equation}
where $k$ is the unique index identified in (\ref{eq:Lemma3,result1}). Next, we define the set of relevant clusters as $\tilde{\mathcal{K}} = \mathcal{K}_{\widetilde{HC} \bowtie (\Delta^{(n)}-\tilde{L})}$, where the symbol $\bowtie$ denotes the cluster intersection relation between $\widetilde{HC}$ and $\Delta^{(n)} - \tilde{L}$, see (\ref{eq:intersect}).

Using this definition, we expand the cluster potential function step-by-step:
\begin{equation}
\begin{aligned}
\omega_{\text{cluster}(\widetilde{HC}, \Delta^{(n)} - \tilde{L})} (\vec{\square}) 
&= \sum_{\phi_{2m} \in \tilde{\mathcal{K}}} \omega_{\phi_{2m}}(\vec{\square}) \\
&= \sum_{\phi_{2m} \in \tilde{\mathcal{K}}} \delta_{c(\vec{\square}), c(\phi_{2m})} 
  \quad (\text{by the definition of $\omega_{\phi_{2m}}$ (\ref{def:potential of cluster})} ) \\
&= \sum_{\phi_{2m} \in \tilde{\mathcal{K}}} \delta_{c(\vec{\square}), c(\vec{\square}_c) + \sum_{i=1}^{2m-1} h s_i} 
  \quad (\text{by the definition of $c(\phi_{2m})$ (\ref{c of cluster})}) \\
&= \sum_{\phi_{2m} \in \tilde{\mathcal{K}}} \delta_{c(\vec{\square}) - c(\vec{\square}_c), \sum_{i=1}^{2m-1} h s_i} 
  \quad (\text{rearranging the delta function})
\end{aligned}
\end{equation}

note that the first term under $\delta$ has 1 largest component while the second has $2m-1$ ($m>2$), so they can't be equal
\begin{equation}
\omega_{\text{cluster}}(\vec{\square}) = 0.
\end{equation}
\end{proof}

\subsection{Lemma 4}
\label{subsec:Lemma4}

 For $\forall$ $\Delta^{(n)}$ and $\vec{\square}\in F(d,\{\vec{e}_{n_i}\})$, $d<n$,
\begin{equation}
\omega_{0,\Delta^{(n)}-\tilde{L}} (\vec{\square}) = 0.
\end{equation}
\begin{proof}
We begin by leveraging the condition \(d < n\) (where \(d\) denotes the dimension of the surface set containing \(\vec{\square}\)). By the definition of \(d\)-dimensional surface points (see Definition of surface set \ref{def:surface}), this dimension condition implies there exists at least one index \(i\) such that the \(i\)-th component of \(\vec{\square}\) is zero, i.e.,
\begin{equation} d < n \Rightarrow \forall j\neq n_i, \quad l_j(\vec{\square}) = 0. \end{equation}

Next, consider an arbitrary vector \(\tilde{\vec{\square}} \in \Delta^{(n)} - \widetilde{L}\) (the remaining partition after removing \(\widetilde{L}\) via the bisect operation). By the definition of L (\ref{def:bisect}), there exists an integer k, such that:\\
\begin{equation}
l_{n_k}(\tilde{\vec{\square}}) - l_{n_k}\left(\vec{\square} - \sum e_{n_i}\right)  < -1.
\end{equation}

We now analyze the component-wise difference \(l_j(\tilde{\vec{\square}}) - l_j(\vec{\square})\):

Substituting \(\vec{\square} = \vec{\square} - \sum e_{n_i} + \sum e_{n_i}\), we find:
\begin{equation} l_{n_k}(\tilde{\vec{\square}}) - l_{n_k}(\vec{\square}) = l_{n_k}(\tilde{\vec{\square}}) - \left( l_{n_k}\left(\vec{\square} - \sum e_{n_i}\right) + 1 \right) < -1\end{equation}

Combining this with the fact that \(l_j(\vec{\square}) = 0\), we further derive the difference between the \(j\)-th and \(n_k\)-th components of \(\tilde{\vec{\square}} - \vec{\square}\):
\begin{equation} l_j(\tilde{\vec{\square}} - \vec{\square}) - l_{n_k}(\tilde{\vec{\square}} - \vec{\square}) > 1 + l_j(\tilde{\vec{\square}}) > 1. \end{equation}

By the property of the neighborhood relation (\ref{eq:neighbor pr2}), this implies the $n-1$ vector \(C(\tilde{\vec{\square}})\) and \(C(\vec{\square})\) are not neighbors, denoted as:
\begin{equation}  C(\tilde{\vec{\square}}) \nleftrightarrow C(\vec{\square}). \end{equation}

The property of neighbors (\ref{eq:neighbor pr1}) gives a key consequence for the potential function: for any subset \(I \subseteq \Delta^{(n)} - \widetilde{L}\) and any \(\vec{\square}' \in I\), removing \(\vec{\square}'\) from \(I\) does not change the potential at \(\vec{\square}\). 
\begin{equation} \forall I \subseteq \Delta^{(n)} - \widetilde{L}, \quad \vec{\square}' \in I, \quad w_{I - \vec{\square}'}(\vec{\square}) = w_I(\vec{\square}). \end{equation}

We can iteratively apply this result by removing all elements from \(\Delta^{(n)} - \widetilde{L}\) one by one. Eventually, we reduce \(I\) to the empty set \(\emptyset\), and since the potential function of the empty set at any position is zero (\(w_{\emptyset}(\vec{\square}) = 0\)), we conclude:
\begin{equation}  w_{0,\Delta^{(n)} - \widetilde{L}}(\vec{\square})=w_{\Delta^{(n)} - \widetilde{L}}(\vec{\square}) = w_{\emptyset}(\vec{\square}) = 0. \end{equation}

\end{proof}
%=============================================================================

\bibliography{XFZCZ}

\end{document}